\documentclass[apj]{emulateapj}

\slugcomment{Accepted for Publication in \emph{The Astrophysical Journal}}
\shorttitle{Extragalactic Zeeman Detections in OH Megamasers}
\shortauthors{Robishaw, Quataert, \& Heiles}

\begin{document}


\title{Extragalactic Zeeman Detections in OH Megamasers}
\author{Timothy Robishaw, Eliot Quataert, \& Carl Heiles}
\affil{Department of Astronomy, University of California at Berkeley, 601 Campbell Hall, Berkeley, CA 94720-3411;
\\robishaw@astro.berkeley.edu, eliot@astro.berkeley.edu, heiles@astro.berkeley.edu}


\begin{abstract}

  We have measured the Zeeman splitting of OH megamaser emission at
  1667 MHz from five (ultra)luminous infrared galaxies ([U]LIRGs)
  using the 305~m Arecibo telescope and the 100~m Green Bank
  Telescope.  Five of eight targeted galaxies show significant
  Zeeman-splitting detections, with 14 individual masing components
  detected and line-of-sight magnetic field strengths ranging from
  $\simeq 0.5-18$ mG.  The detected field strengths are similar to those
  measured in Galactic OH masers, suggesting that the {\it local}
  process of massive star formation occurs under similar conditions in
  (U)LIRGs and the Galaxy, in spite of the vastly different
  large-scale environments.  Our measured field strengths are also
  similar to magnetic field strengths in (U)LIRGs inferred from
  synchrotron observations, implying that milligauss magnetic fields
  likely pervade most phases of the interstellar medium in (U)LIRGs.
  These results provide a promising new tool for probing the
  astrophysics of distant galaxies.

\end{abstract}


\keywords{galaxies: magnetic fields --- ISM: magnetic fields --- magnetic
  fields --- masers --- polarization --- radio lines: galaxies}




\section{Introduction}
\label{sec:intro}

(Ultra)luminous infrared galaxies ([U]LIRGs) are a population of galaxies
that emit far-infrared (FIR) radiation with energies comparable to those of
the most luminous quasars ($\log(L_{\rm FIR}/L_\odot)>11$ and 12 for LIRGs
and ULIRGs, respectively).  Nearly every ULIRG appears to have undergone a
merger/interaction and contains massive star formation and/or an active
galactic nucleus (AGN) induced by gravitational interactions.  \citet{lo05}
details very long baseline interferometry (VLBI) observations of the 1667
MHz hydroxyl (OH) transition in the nuclear regions in (U)LIRGs that have
revealed multiple masing regions with $1<\log(L_{\rm OH}/L_\odot)<4$; these
regions are known as OH megamasers (OHMs).  Each OHM has a spectral
linewidth of between 50 and 150 km s$^{-1}$; when viewed by a single dish,
these spectral components are superimposed.  The 1667 MHz OHM flux density
is always a few to many times that of the 1665 MHz transition, and in many
cases the 1665 MHz line is absent \citep{darlingg02}; this is an
interesting contrast to the case of OH masers in the Galaxy in which the
1665 MHz transition is usually dominant \citep{reidm88}.  The starbursts
and AGNs in ULIRGs create strong FIR dust emission as well as a strong
radio continuum; the OHMs are generally believed to be pumped by the FIR
radiation field \citep[e.g.,][]{randellfjyg95} although collisional
excitation may be important as well \citep[e.g.,][]{lonsdalelds98}.
\citet{lockette08} have recently suggested that the 53 $\mu$m OH pump lines
in addition to line overlap of large ($\gtrsim$ 10 km s$^{-1}$) turbulent
linewidths can account for the observed dominance of the 1667 MHz
transition in OHMs.  They further argue that pumping due to FIR radiation
can explain all observed main-line OH masers, both those in Galactic
star-forming regions and those in OHM galaxies.  Given the conditions that
exist in ULIRGs and considering that so many OH masers in our Galaxy are
associated with massive star-forming regions \citep{fishram03}, it is
therefore not surprising that the entire OHM sample finds homes in LIRGs,
strongly favoring the most FIR-luminous, the ULIRGs \citep{darlingg02}.

The high gas and energy densities in ULIRGs make them natural locations to
expect very strong magnetic fields.  Much of the radio emission in ULIRGs
is resolved on scales of $\sim 100$ pc with VLA observations
\citep{condonhyt91}.  High-resolution observations of Arp 220
\citep{rovilosdlls03} show that the OHMs arise in this region as well.
With this size scale and the observed radio flux densities, minimum energy
arguments suggest {\it volume averaged} field strengths of $\approx$ 1 mG
\citep[e.g.,][]{condonhyt91,thompsonqwmm06}, which are significantly larger
than the $\sim 10~\mu$G fields in normal spirals.  The field strengths in
ULIRGs cannot be much below a mG or else inverse Compton cooling would
dominate over synchrotron cooling, making it energetically difficult to
explain the radio flux densities from ULIRGs and the fact that ULIRGs lie
on the FIR-radio correlation. The field strengths could, however, in
principle be larger than the minimimum energy estimate if, as in our
Galaxy, the magnetic energy density is in approximate equipartition with
the total pressure \citep{thompsonqwmm06}.  The latter can be estimated
from the observed surface density.  CO observations of Arp 220 and several
other systems reveal $\sim 10^9 M_\odot$ of molecular gas in the central
$\sim 100$ pc \citep[e.g.,][]{downess98} implying gas surface densities
$\Sigma \sim 1-10$ g cm$^{-2}$, $10^3-10^4$ times larger than in the Milky
Way (MW).  The equipartition field scales as $B \propto \Sigma$, implying
that the mean field in ULIRGs could approach $\sim 10$ mG.

Motivated by the above considerations, we carried out a survey of 8
(U)LIRGs searching for Zeeman splitting in OHMs.  This paper presents our
results, which represent the first detections of extragalactic Zeeman
splitting from an emission line and the first extragalactic detections
within an external galaxy proper.  The only previous extragalactic
detection was made by \citet{kazestc91} and confirmed by \citet{sarmamtc05}
via absorption of 21~cm emission in a high-velocity system towards NGC 1275
(Per A).  \S~\ref{sec:sources} outlines what is known about each of our
targets.  In \S~\ref{sec:obs} we describe the observations.  In
\S~\ref{sec:reduction} we discuss the data reduction and calibration
method.  \S\S~\ref{sec:results} and \ref{sec:discussion} present a summary
and discussion of the results, respectively.


\section{Source Selection}
\label{sec:sources}

In selecting our sample of targets from the compilation of all known OHMs
by \citet{darlingg00,darlingg01,darlingg02}, we chose the three simplest
criteria possible.  We selected 12 (U)LIRGs: ({\it a}) with the largest OHM
peak flux densities; ({\it b}) whose discoverers did not regard the OHM
detection validity as suspicious; ({\it c}) that are observable from
Arecibo, Puerto Rico, or Green Bank, West Virginia.  Our sample includes 2
of only 3 known OH gigamasers \citep[$L_{OH} > 10^{4}
L_{\odot}$;][]{darlingg02}.  Here we summarize what is known about each
source and its OHM emission.\footnote{Unfortunately, we only obtained
  usable data for 2 of the 6 sources observed at Green Bank, therefore we
  only provide source descriptions for the 8 (U)LIRGs for which we have
  presentable results.}

\noindent{\bf IRAS F01417$+$1651}: This LIRG is most commonly
known\footnote{As a shorthand, when referring to source names in the text
  by their IRAS designation, we shall henceforth use the right ascension
  designator only; we shall refer to 01417, 10038, and 15327 by their more
  common designators, III Zw 35, IC 2545, and Arp 220, respectively.  We
  shall retain the full IRAS designation in figure and table captions as
  well as section headings.} as \object[III Zw 35]{III Zw 35} and has an
optical heliocentric redshift of $z = 0.0274$.  It is a double galaxy
system and is classified as a Seyfert~2 galaxy.
\citet{staveley-smithccpu87} present a single-dish spectrum from the
Jodrell Bank Mk1A 76~m telescope showing emission from the 1667 MHz
transition at a velocity of 8262 km s$^{-1}$ with a peak flux density of
240 mJy and a total velocity extent of 270 km s$^{-1}$ at the 10\% flux
density level.  The line profile can easily be seen to have at least three
components.  The 1665 MHz line is also weakly detected ($\simeq 25$ mJy) and
completely separated from the 1667 MHz emission, with an estimated
hyperfine line ratio (defined as $R_{H} \equiv \int f_{1667}\,d\nu / \int
f_{1665}\,d\nu$, where the integrals represent the total flux density of
each transition) of $R_{H} \approx 9$.

\citet{killeensws96} observed III Zw 35 using the Australia Telescope
Compact Array (ATCA).  Their goal was an attempted detection of Zeeman
splitting in the OHM emission.  The peak flux density was 247.6 mJy and
their sensitivity was 3.6 mJy.  They observed no Zeeman splitting and their
model-dependent estimate for a 3$\sigma$ upper limit on the line-of-sight
magnetic field was 4.0 mG.

\citet{diamondlls99} present global VLBI observations of the OHM emission
in III Zw 35.  They label two regions of 1667 MHz OHM emission in the south
(S1 and S2) and three in the north (N1--N3), each region covering about 20
mas and separated by 90 mas.  They recover 60\% of the single-dish flux
density.  \citet{pihlstromcbdp01} performed simultaneous high-resolution
observations of the OHM emission in III Zw 35 using both the European VLBI
Network (EVN; baselines between 198--2280 km) and the Multi-Element
Radio-Linked Interferometer Network (MERLIN; operated by Jodrell Bank
Observatory with baselines between 6.2--217 km).  The map of the 1667 MHz
emission shows two compact regions coincident with the northern and
southern sources of \citet{diamondlls99} connected by two bridges of
weaker, more diffuse emission.  In total, 80\% of the single-dish flux
density was recovered.  A velocity gradient of $\simeq 1.5$ km s$^{-1}$
pc$^{-1}$ is observed from the southern to the northern regions and is
evident in the diffuse component.  The emission is modelled as a torus of
multiple maser clouds inclined at 60$^{\circ}$; the compact OHM emission
would be seen at the tangent points where a few clouds could be
superimposed in such a fashion that strong OHM emission would be produced
from the foreground clouds amplifying those in the background.  At the
front and back of the torus, the emission would be weak because the path
lengths through the torus are small and the clouds are less likely to
overlap \citep{pihlstromcbdp01,parracep05}.

\noindent{\bf IRAS F10038$-$3338}: Also known as \object[IC 2545]{IC 2545},
this LIRG is a set of interacting galaxies at $z = 0.0341$.  A single-dish
spectrum made using the Parkes 64~m telescope is presented by
\citet{staveley-smithncawr92} showing 1667 MHz OHM emission centered at
10093 km s$^{-1}$ with a full width at half maximum (FWHM) velocity range
of 63 km s$^{-1}$ and a peak flux density of 315 mJy.  Likely due to its
low declination, there have been no VLBI observations of this source
despite its brightness.

\citet{killeensws96} present an ATCA spectrum with much better sensitivity
(5.4 mJy in Stokes $I$) and velocity resolution than that of
\citet{staveley-smithncawr92}; the 1667 MHz emission contains five narrow
peaks superimposed on a broad emission component.  The brightest component
has a peak flux density of 260 mJy and a velocity of 10097 km s$^{-1}$.
\citet{killeensws96} failed to detect Zeeman splitting and estimated that
the line-of-sight field should be less than 4.3 mG.

\noindent{\bf IRAS F10173$+$0829}: The only single-dish observations of the
OHM line emission in this LIRG at $z = 0.0480$ were made with the 305~m
Arecibo telescope and are detailed in \citet{mirabels87}. There are two
distinct peaks in the profile of the 1667 MHz emission with a separation of
about 100 km s$^{-1}$, the dominant peak having a velocity of 14720 km\
s$^{-1}$ with a FWHM of 39 km s$^{-1}$ and a peak flux density of about 105
mJy.  The 1665 and 1667 MHz lines are well separated with a hyperfine line
ratio $R_{H} = 14.6$.

MERLIN observations made by \citet{yu04,yu05} show roughly 50 maser spots
distributed into three clumps, labelled east, central, and west, over an
area of $1\farcm4 \times 0\farcm6$.  The spots within each clump are
distributed along a line, with each of the three lines having a different
direction; \citet{yu05} proposes (without much justification) that the
spots may be distributed along a warped circumnuclear torus seen edge-on.
The OHM emission is seen only at the 1667 MHz transition and is coincident
with the infrared central position.

\noindent{\bf IRAS F11506$-$3851}: Also known as ESO 320-30, this LIRG is
classified as an \ion{H}{2} galaxy at $z = 0.0108$.  A Parkes single-dish
spectrum is presented by \citet{staveley-smithncawr92} showing 1667 MHz OHM
emission centered at 3103 km s$^{-1}$ with a FWHM velocity extent of 87 km
s$^{-1}$ and a peak flux density of 105~mJy. There is neither enough
sensitivity nor bandwidth to clearly discern any 1665 MHz emission.  There
are no interferometric observations of this source: like IC 2545, the low
declination of 11506 would hinder any attempted VLBI observations.

\noindent{\bf IRAS F12032$+$1707}: A gigamaser discovered at Arecibo by
\citet{darlingg01}.  The host object, a ULIRG at $z = 0.2170$, has been
classified as a LINER-type AGN \citep{veilleuxks99}.  The OHM emission
spans almost 2000 km\ s$^{-1}$ with a redshifted high-velocity tail and a
mean flux density of roughly 9 mJy. The 1665 and 1667 MHz lines are
impossible to distinguish and clearly blended.  A very narrow and
bright component is seen at 64500 km s$^{-1}$ with a peak flux density of
16.3 mJy.

\citet{pihlstrombdk05} used the Very Long Baseline Array (VLBA) to show
that the OHM emission is confined to an area of $25 \times 25$ mas.  All
the single-dish flux density was recovered. They were able to clearly
identify five peaks in their Stokes $I$ spectrum that corresponded with
Darling's single-dish spectrum.  By averaging channels around each peak,
they found the maser components were spatially separated and aligned
roughly north-south, implying an ordered velocity gradient. No continuum
emission was detected, implying that the continuum emission is resolved out
on scales less than 75 mas.

\noindent{\bf IRAS F12112$+$0305}: This ULIRG is classified as a LINER-type
AGN and is an interacting pair of galaxies at $z = 0.0730$.  The only
information concerning the OHM emission in this ULIRG is listed in tabular
form in \citet{staveley-smithncawr92}; no spectrum has been published.  The
1667 MHz line was measured at a velocity of 5540 km\ s$^{-1}$ and no
information about the 1665 MHz transition is published.  Its 1667 MHz flux
density is listed in \citet{darlingg02} as 45 mJy. There are no VLBI
observations of the 1667 MHz OHM emission for this source.

\noindent{\bf IRAS F14070$+$0525}: Discovered by \citet{baanrfah92}, this
gigamaser is the most distant OHM at a redshift of $z = 0.2644$.
\citet{darlingg02} redetected this source in their survey.  The OH lines
are so wide and blended (1580 km\ s$^{-1}$ at 10\% peak flux density) that
it is impossible to identify any 1665 MHz emission. The spectral line
profile measured by \citet{darlingg02} has a peak flux density of 8.4 mJy
and shows no significant changes since the original detection by
\citet{baanrfah92}.  \citet{darlingg02} suggest that many peaks in the
profile are likely the result of many masing nuclei within the host ULIRG,
which is classified as a Seyfert~2.

VLBA observations made by \citet{pihlstrombdk05} recovered less than 10\%
of the single-dish flux density and only two of the many single-dish
spectral peaks were detected.  Most of the single-dish emission is
therefore diffuse.  The spatial extent of the VLBA emission is confined to
$10 \times 10$ mas.

\noindent{\bf IRAS F15327$+$2340}: This is perhaps the most well-known
ULIRG and is better known as \object[Arp 220]{Arp 220} or IC 4553.
\citet{baanwh82} discovered the OHM emission using Arecibo and list the
single-dish properties of the OHM emission as having a velocity of 5375 km\
s$^{-1}$ and FWHM velocity extent of 108 km s$^{-1}$.  The spectrum clearly
shows that the 1665 and 1667 MHz transitions are distinct with a hyperfine
line ratio $R_{H} = 4.2$ and a peak flux density of 320 mJy.


\citet{smithlld98} used global VLBI continuum imaging at 18~cm to show that
the high brightness temperature core of Arp 220 is composed of multiple
compact sources, which they interpret as luminous radio supernovae (RSNe).
These RSNe are not coincident with the compact 1667 MHz OHM spots
discovered by \citet{lonsdalelds98}. More recently, \citet{lonsdaledtsl06}
have used high-sensitivity 18 cm VLBI observations of the nuclei to detect
4 previously unseen sources in a one year period, supporting the RSNe
interpretation.  \citet{parracdtlls07} have made the first multiwavelength
observations of these compact sources; they identify a fraction of these
sources to be supernova remnants.

\citet{rovilosdlls03} present MERLIN maps of the 18~cm continuum and OHM
line emission in Arp 220; two components are seen roughly 1\arcsec\ apart,
each coinciding with a nucleus imaged in the infrared by
\citet{grahamcmnsw90}.  The OHM emission is resolved into one component
aligned with the eastern continuum feature and two components that are
aligned north to south straddling the western nucleus.
\citet{lonsdalelds98} and \citet{rovilosdlls03} present global VLBI
spectral line maps that show that the OHM emission is resolved into
multiple compact spots.  The northernmost features in both the eastern and
western nuclei form elongated ridges.


\section{Observations}
\label{sec:obs}

In 2006 February we used the $L$-band wide receiver of the 305~m
Arecibo\footnote{The Arecibo Observatory is part of the National Astronomy
  and Ionosphere Center, which is operated by Cornell University under a
  cooperative agreement with the National Science Foundation.} telescope in
full-Stokes mode in an attempt to detect Zeeman splitting of the 1667 MHz
OH transition in the 6 positive-declination sources listed in
\S~\ref{sec:sources}.

Since the spatial extent of each source is much smaller than Arecibo's
$3\farcm3$ beam, our observing method was to simply spend equal time at
on-source and off-source positions.  In this position-switching scheme, we
alternated between 4 minutes on source and 4 minutes at a reference
position having the same declination as the source and a right ascension 4
minutes east of the source.  In this way, the hour angle ranges of the
source and reference observations were nearly identical.  Our integration
time was one second, allowing us to remove short-term radio-frequency
interference (RFI).  The total integration time for each source is as
follows: 5.4 hours for III Zw 35, 2.7 hours each for 10173 and 12032, 3.1
hours for 12112, 4.6 hours for 14070, and 5.9 hours for Arp 220.  We
configured the correlator to produce 4 spectra per integration: one 6.25
MHz bandpass centered on the mean of the 1665 and 1667 MHz transitions; 2
narrow bandwidths (either 3.125 or 12.5 MHz, depending on the velocity
extent of the source) centered on the 1665 and 1667 MHz transitions,
respectively; and one wide bandwidth (either 12.5 or 25 MHz) centered on
the mean of the 1665 and 1667 MHz transitions. We calibrated the Mueller
matrix for Stokes parameters using the standard Arecibo technique
\citep{heilespnlbgloetal01,heilest04} of observing spider scans on the
linearly-polarized continuum sources 3C 138 and 3C 286.

In 2005 December we used the $L$-band receiver of the 100~m Robert C.\ Byrd
Green Bank Telescope\footnote{The National Radio Astronomy Observatory is a
  facility of the National Science Foundation operated under cooperative
  agreement by Associated Universities, Inc.} (GBT) to observe an
additional six OHM galaxies. All but the two sources in
\S~\ref{sec:sources} at negative declinations were affected by insidious
RFI that left our data corrupted beyond salvation (For completeness, the
sources that were obliterated by RFI were IRAS F12540$+$5708, IRAS
F13428$+$5608, IRAS F17207$-$0014, and IRAS F20100$-$4156.) We used two
observing methods for each source: position switching (as described above)
and least-squares frequency switching (LSFS; see \citealt{heiles07} for the
details of this observing method and its corresponding reduction scheme).
The LSFS method was used to accurately derive the gain for each
integration; the data were then combined in the standard way using the
off-source, position-switched spectra.  Our off-source positions were 23
minutes east of each on-source position in order to cause the GBT to track
as closely as possible the path of our on-source observations.  We used a
12.5 MHz bandwidth and 9-level sampling for all observations.

We accumulated 4.0 hours of RFI-free, on-source integration time for IC
2545 and 5.8 hours for 11506.  As we did at Arecibo, we
observed 3C 286 using spider scans in order to calibrate the $L$-band
Mueller matrix at the GBT.


\section{Data Reduction}
\label{sec:reduction}

The complex Stokes $I$ (see footnote\footnote{We use the classical
  definition of Stokes $I$, which is the sum (not the average) of two
  orthogonal polarizations.  Thus, stated Stokes $I$ flux densities are
  twice those listed in \S~\ref{sec:sources} and in other catalogs.})  line
shape in each of the maser sources is a composite of many narrow maser
lines at various velocities spread about the systemic velocity of the
system.  Therefore, we chose to least-squares fit each line profile with
multiple Gaussian components. Without VLBI observations, it is impossible
to attribute any particular velocity or width to a Gaussian component
within the profile.  The only method available to us for assessing a
possible field strength from each Stokes $V$ spectrum was to decompose each
$I$ profile into the fewest number of Gaussian components that would yield
reasonable residuals while also allowing enough components to reproduce the
multiple splittings in the $V$ spectrum.  None of the parameters in our
multiple-component Gaussian fits were held fixed.  We discuss our method in
more detail in \S~\ref{sec:gauss}.

\subsection{Calibration}
\label{sec:calibration}

The derived Mueller matrix was applied to all OHM observations to correct
the polarization products and obtain the pure Stokes spectra for each
observed source.  We converted from antenna temperature to flux density by
assuming the antenna gain to be 10.0 K Jy$^{-1}$ at Arecibo and 2.0 K
Jy$^{-1}$ at the GBT; these gains were estimated from observations of
standard flux density calibrators.

We follow the IAU definition for Stokes $V$, namely $V = RHCP-LHCP$, where
$RHCP$ is the IEEE definition of right-hand circular
polarization.\footnote{Defined as a clockwise rotation of the electric
  vector along the direction of propagation.}  We determined the sense of
Stokes $V$ at Arecibo by observing the highly circularly polarized 1665 MHz
Galactic maser W49(OH); the result is consistent with the measurements of
\citet{rogersmcbmbh67} and \citet{colesr70}.  The sense of Stokes $V$ has
not yet been determined for the GBT Autocorrelation Spectrometer, which
only began functioning with full-Stokes capability months prior to these
observations.

\subsection{RFI Removal}
\label{sec:rfi}

We examined each set of spectra, both off source and on, for RFI and
rejected suspicious-looking data, which constituted only a few percent for
only two sources, III Zw 35 and Arp 220. The other sources were completely
free of RFI except for occasional monochromatic signals whose topocentric
frequencies are constant; fortunately, most of these fall off the OHM
lines.

For one source, 10173, the monochromatic RFI fell on the OHM line. We
observed this source over several days, during which the changing Doppler
shift moved the RFI across part of the OH spectrum. For each day the RFI
was a sharp spike with the usual ringing sidelobes. We Hanning smoothed
each spectrum, which eliminated the ringing, and interpolated across each
day's spike, which effectively removed the RFI.

The source III Zw 35, whose OH lines are centered near 1622.5 MHz, was
highly contaminated by RFI that probably arises from the Iridium
communications satellites. The RFI consists of a spikey pattern that
repeats periodically across the spectrum at about a 0.33 MHz interval. It
was impossible to obtain reasonable results by averaging data. However, by
taking medians instead of averaging, the spectra look quite good and the
RFI is reduced to levels of about 20 mJy in Stokes $I$ and 3 mJy in Stokes
$V$ --- levels that are considerably smaller ($<1\%$) than the OHM spectral
features.

\subsection{Bandpass and Gain Correction}
\label{sec:bandpass_and_gain}

We correct our spectra for the intermediate-frequency bandpass. Both the
Arecibo $L$-band wide receiver and the GBT $L$-band receiver are
dual-polarized feeds with native linear polarization.  For Stokes $I$ and
$Q$, we divide each of the two linear polarization spectra ($XX$ and $YY$)
by its associated off-source spectrum; then we add the results to obtain
Stokes $I$ and subtract them to obtain $Q$. To generate Stokes $U$ and $V$,
we combine the cross-correlation spectra ($XY$ and $YX$) having divided by
the square root of the product of the off-source $XX$ and $YY$ spectra.

We always show difference spectra: on-source minus off-source. Normally,
on- and off-source spectra are combined by subtracting the latter from the
former.  If the two spectra have equal noise $\sigma$, then the noise in
the difference is $\sqrt{2}\sigma$. Our off-source spectra have no
fine-scale frequency structure, so we can reduce the noise by smoothing. We
use a Fourier technique to smooth the off-source average spectrum.  By
zeroing lags at high delays in the autocorrelation function of the average
off-source spectrum and then Fourier transforming, we nearly eliminate the
noise contribution from the off-source spectrum while retaining the shape
of the bandpass.  This reduction in noise is particularly important for the
polarized Stokes parameters, which are weak.

As mentioned above, Stokes $U$ and $V$ are obtained via the
cross-correlation products $XY$ and $YX$: this insulates them from system
gain fluctuations. However, since Stokes $Q$ is the difference between the
two native linear polarizations, it is susceptible to gain fluctuations.
We defer the discussion of linear polarization to \S~\ref{sec:linear_pol}.

After gain and bandpass correction, the on- and off-source spectra were
averaged separately and combined to yield the final average Stokes spectra.

\subsection{Fitting Gaussian Components to Stokes $I$ Profiles}
\label{sec:gauss}

Fitting multiple Gaussians to complicated spectral profiles carries a
significant degree of subjectivity because the fits are nonlinear.
Generally, nonlinear fits require beginning from initial ``guessed''
parameters and letting the fit converge with successive iterations
\citep{presstvf92}. Nonlinear fits usually have multiple $\chi^{2}$ minima
and the particular minimum selected depends on the initial guesses, which
in turn depend on the subjective judgement of the person doing the fitting.
Therefore, we outline the following guidelines that were used for selecting
the initial guesses for Gaussian components in the fits to the Stokes $I$
spectra:
\begin{enumerate}

\item For each peak (i.e., local maximum) in $I$, we included a single
  Gaussian component whose three parameters (flux density, central
  frequency, and frequency width) were visually estimated.

\item Many $I$ peaks are distinctly asymmetric. We fitted these asymmetries
  by including one or two Gaussian components with visually-estimated
  parameters in addition to the central component of guideline (1).

\item The Gausssian components estimated in guidelines (1) and (2) are
  usually fairly narrow and lie on top of one or two underlying broader
  lines --- core-halo structure.  We included one or two broad Gaussian
  components to represent these broader lines.

\item With all of the above, our goal was to use the fewest number of
  Gaussian components that would yield reasonable residuals.

\end{enumerate}

It was straightfoward to apply the above guidelines to the sources III Zw
35, IC 2545, 11506, 12112, 14070, and even Arp 220, for which we fitted 18
components.  The sources 10173 and 12032 are somewhat more complex.  In
\S\S~\ref{sec:v_014017}--\ref{sec:v_15327} we describe how we applied the
above selection guidelines when appropriate.

We stress that our Gaussian-component representations are not unique.
Particular problems include:
\begin{enumerate}

\item In guideline (1) above, noise prevents us from identifying weak
  components.  This introduces a sensitivity cutoff.  Noise also prevents
  us from distinguishing two or more closely-spaced blended real components
  from a single broader component.  Because we favor choices with the
  fewest number of components, this introduces a bias towards wider
  components.

\item In guidelines (2) and (3) above, whether to represent a peak needing
  multiple Gaussians by an asymmetry or core-halo structure can be
  extremely subjective.  For example, the combination of two narrow
  Gaussians separated by a fraction of their FWHMs can closely mimic the
  combination of a broad and a narrow Gaussian with roughly the same
  centers.

\end{enumerate}

In summary, for most sources our Gaussian fits follow our fitting
guidelines in a reasonably straightforward fashion; if the fits were done
by other people who followed these guidelines, the components would be
mostly reproduced. However, Gaussian component fitting has uncertainties as
mentioned above, particularly when components are blended and
signal-to-noise (S/N) is low.


\section{Results}
\label{sec:results}

For each source, the Stokes $I$ spectrum exhibits a fairly broad,
relatively smooth underlying profile for the OHM emission on top of which
small bumps from individual masers can be seen.  VLBI studies show that the
underlying profile arises from spatially extended OH emission and,
sometimes, an assembly of many masers that are not individually
recognizable \citep{pihlstrombdk05,diamondlls99}.  We fit the Stokes $I$
spectrum for each source with a series of Gaussian profiles.  For Arp 220,
12112, and 12032 we also fit a first-degree polynomial (12032 required a
second-degree polynomial in addition), since these profiles exhibit broad
wings.  A few sources have a large number of discernable individual masers:
for example, we used 18 Gaussians components for Arp 220.

We examined circular polarization for each source, and linear polarization
for the six Arecibo sources only.  Five of the eight sources exhibit
significant circular polarization that is interpretable as Zeeman
splitting, particularly for the recognizable individual maser components.
For four of the sources, there is evidence that the magnetic field reverses
direction between OHM spots within the source.

We see detectable linear maser polarization in two sources (possibly four)
and are able to estimate Faraday rotation in both. We present all spectra
as a function of frequency as viewed in the heliocentric frame.  Since all
OHMs are extragalactic sources, OHM spectra are almost always presented
versus optical heliocentric velocity $v_{\odot}$, which is conventionally
defined as
\begin{equation}
{v_{\odot} \over c} \equiv {\nu_{0} \over \nu} - 1 \equiv z_{\odot}\ ,
\end{equation}
where $c$ is the speed of light, $\nu_{0}$ is the rest frequency (which
is taken to be 1667.359 MHz for OHMs since this transition always dominates
the 1665.4018 MHz transition), $\nu$ is the observed frequency, and
$z_{\odot}$ is the redshift of the maser.

First, in \S~\ref{sec:circular_pol} we present the circular polarization
results for each source in addition to describing the total intensity
properties.  We present the linear polarization results for each source
observed with Arecibo in \S~\ref{sec:linear_pol}.  For sources in which the
1665 and 1667 MHz emission are separable, we calculate the hyperfine ratio
$R_{H}$.


\subsection{Circular Polarization and Line-of-Sight Magnetic Fields}
\label{sec:circular_pol}

For the usual case in which the Zeeman splitting is small compared to the
linewidth, the Stokes $V$ spectrum is given by
\begin{equation} \label{eq:v}
V = \left({\nu \over \nu_{0}}\right) \left({dI \over d\nu}\right) b B_{\parallel}\ ,
\end{equation}
where $B_{\parallel}$ is the line-of-sight component of the magnetic field
at the OHM and $b$ is known as the {\em splitting coefficient}\footnote{The
  splitting coefficient is directly proportional to the Land\'e $g$-factor
  for the transition: $b = 2g\mu_{0}/h$, where $\mu_{0}$ is the Bohr
  magneton and $h$ is Planck's constant.} \citep{heilesgmz93}, equal to
1.96 Hz $\mu$G$^{-1}$ for the OH 1667.359 MHz
transition;\footnote{\citet{modjazmkg05} made a valiant effort to detect
  Zeeman splitting of 22.2 GHz H$_{2}$O megamasers in NGC 4258 using the
  VLA and the GBT, but the splitting coefficient for this hyperfine
  transition is nearly 1000 times weaker than that of the 1667 MHz OH
  transition.} the factor ${\nu/\nu_{0}}$, equivalent to
$(1+z_{\odot})^{-1}$, accounts for the frequency compression of redshifted
lines.  In order to derive a magnetic field strength, we need to
least-squares fit the Stokes $V$ spectrum with the functional form of
equation (\ref{eq:v}).  As is the custom in radio Zeeman work, we add a
term on the right that is linear in Stokes $I$ to account for leakage of
$I$ into the measured $V$.  For $B_{\parallel}>0$ (by convention a positive
magnetic field points away from the observer), if Stokes $V$ is plotted as
a function of frequency, $V$ will be positive on the low-frequency side of
a Stokes $I$ emission line.

We solved equation (\ref{eq:v}) in two ways. In one, we simultaneously
fitted Stokes $V$ for multiple Gaussian components (selected as outlined in
\S~\ref{sec:gauss}) to derive separate, independent magnetic fields for
each Gaussian. In the other, we chose a limited range in frequency $\nu$,
either 0.1 or 0.25 MHz, and fitted for $B_{\parallel}$ for the center of
this range, positioning the center sequentially at each spectral channel to
obtain $B_{\parallel}$ as a function of frequency; we refer to this as the
{\it $B(\nu)$ fit}. The former method is appropriate for individual masers,
while the latter is more suitable for the broad component. We plot the
$B(\nu)$ results only for sources for which the results provide additional
insight.

Performing these fits requires the calculation of $dI/d\nu$. The $I$
profiles are somewhat noisy and the frequency derivative is often very
noisy. This means that traditional least-squares fitting cannot be used,
because it assumes no error in the independent variables.
\citet{saultkzl90} discuss this and suggest using a generalized maximum
likelihood technique. We choose the much simpler approach of using our
multiple-Gaussian fit to the Stokes $I$ spectrum as the independent
variable: it has no noise, so it satisfies the requirements of the
conventional method of least squares.

We present two vertically stacked plots for each source (Figures
\ref{fig:iv_01417}--\ref{fig:iv_15327}).  In the top panel, the Stokes $I$
spectrum is plotted as a solid line and the profiles of the Gaussian
components are plotted as dashed lines.  The residuals (the difference
between the data and the composite Gaussian fit) are plotted with enhanced
vertical scale as a solid line near the middle height of the panel.  Scale
bars are plotted on both the residuals and the baseline of the spectrum to
the right of the OHM emission; the height of each scale bar corresponds to
the labelled flux density.  The bottom axes of both plots show heliocentric
frequency and the top axis of the top plot displays the optical
heliocentric velocity.  In the bottom panel, the Stokes $V$ spectrum is
plotted as a solid line and the dashed line represents the best fit to
equation (\ref{eq:v}).  The integers located between the top and bottom
panels label the number of each Gaussian component as assigned in the
corresponding tabular summary and are positioned at the central frequencies
of each component.\footnote{Where multiple labels overlap, the font size
  has been reduced and the labels stacked corresponding to their associated
  flux densities.}  The displayed spectra and residuals are smoothed over 7
channels for every source.

We assume that each Gaussian represents an emission component for the 1667
MHz transition.  Arp 220 and 12032 have line profiles that are too complex
for the 1665 and 1667 MHz lines to be distinguished.  This introduces some
uncertainty in our Zeeman splitting interpretations in
\S\S~\ref{sec:v_12032} \& \ref{sec:v_15327}. For sources where the 1665 MHz
transition is not blended with the 1667 MHz emission, we present the
spectra showing both transitions in \S~\ref{sec:linear_pol} and calculate
the hyperfine line ratios.  In all cases the 1665 MHz transition was too
weak for Zeeman splitting to be detected even if observed in the 1667 MHz
line.

\subsubsection{IRAS F01417$+$1651  (III Zw 35)}
\label{sec:v_014017}

\begin{figure*}[ht!]
  \begin{center}
    \includegraphics[width=3.5in] {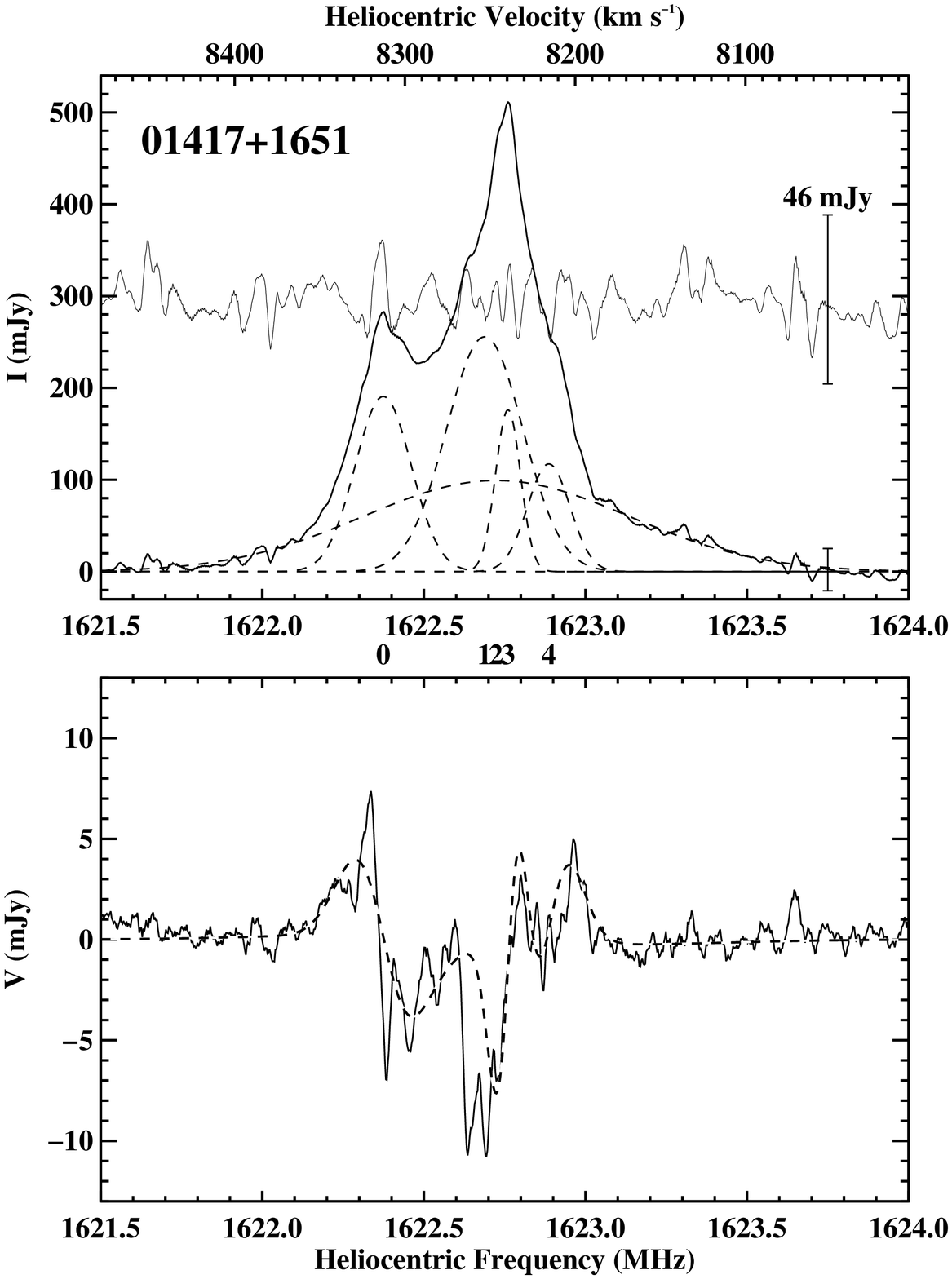}
    \includegraphics[width=3.5in] {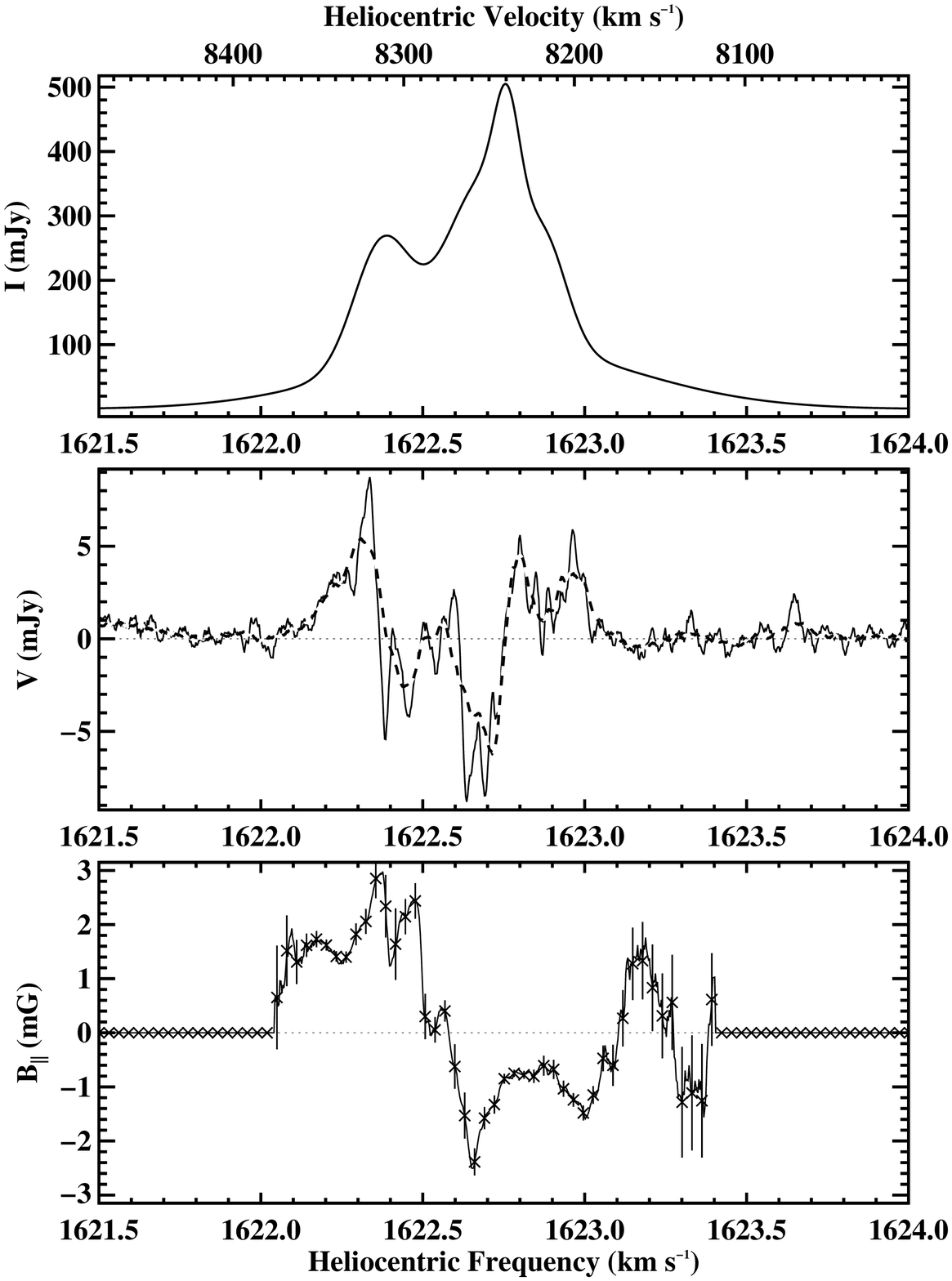}
  \end{center}
  \caption[Total intensity and circular polarization of IRAS F01417$+$1651
  (III Zw 35)] {Total intensity and circular polarization results for IRAS
    F01417$+$1651 (III Zw 35).  ({\it Top left}) The Stokes $I$ spectrum
    ({\it solid line}; twice the conventionally defined flux density) is
    plotted as a function of heliocentric frequency ({\it bottom axis}) and
    optical heliocentric velocity ({\it top axis}).  The profile of each
    Gaussian component is plotted as a dashed line with the corresponding
    component number shown below the frequency axis at the corresponding
    central frequency. Residuals from the composite Gaussian fit are
    plotted through the center of the panel ({\it solid, light-weight
      line}) and are expanded by a factor of 4. The scale bars near the
    right edge of the plot correspond to the labelled flux density range.
    ({\it Bottom left}) The Stokes $V$ spectrum ({\it solid line}) and its
    fit ({\it dashed line}). ({\it Top right}) Composite Gaussian fit to
    Stokes $I$. ({\it Middle right}) Measured Stokes $V$ ({\it solid
      line}); hypothetical Stokes $V$ ({\it dashed line}) produced by a
    uniform $B_{\parallel} = 1$ mG using the derivative of the composite
    $I$ profile above. ({\it Bottom right}) Derived $B_{\parallel}$ ({\it
      crosses and solid line}) and uncertainty ({\it error bars}) from the
    $B(\nu)$ fit.  All spectra and residuals are smoothed with a boxcar of 7
    channels.}
  \label{fig:iv_01417}
\end{figure*}

\begin{figure*}[h!]
\begin{center}
  \includegraphics[width=7in] {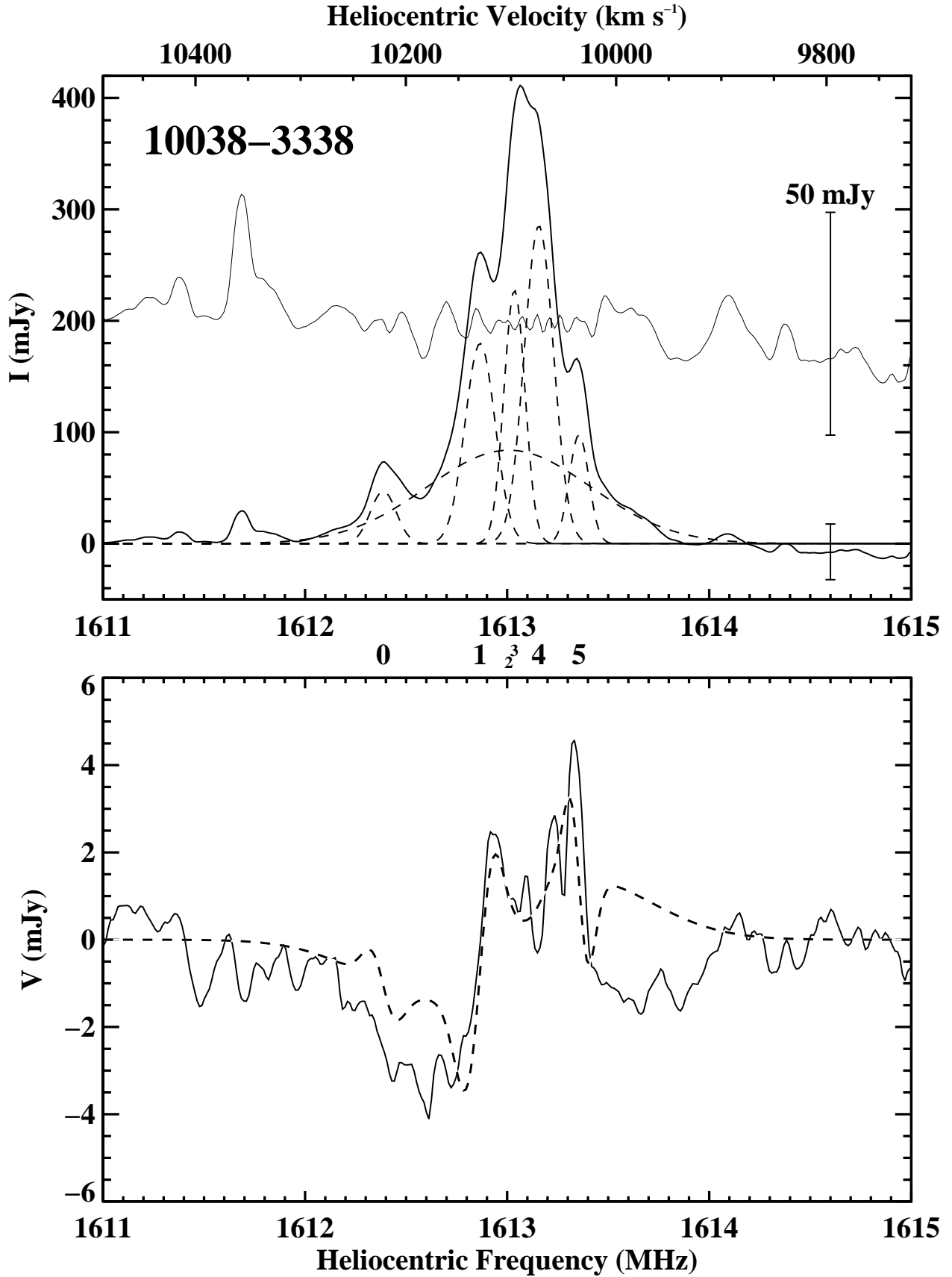}
\end{center}
\caption[Total intensity and circular polarization of IRAS F10038$-$3338
(IC 2545)]{Total intensity and circular polarization results for IRAS
  F10038$-$3338 (IC 2545). See caption for Figure \ref{fig:iv_01417}.
  ({\it Top}) Residuals are expanded by a factor of 4.}
  \label{fig:iv_10038}
\end{figure*}

As we mentioned in \S~\ref{sec:rfi}, our observations of III Zw 35 suffered
severe RFI that we were able to greatly reduce by combining the data using
medians instead of averaging. There remains a spikey pattern that repeats
periodically across the spectrum at an interval of $\simeq$ 0.33 MHz.
Remarkably, this spikey pattern is restricted to an 8 MHz wide interval
centered almost exactly on the OHM lines.  The spikey pattern appears in
both the on-source and off-source spectra, so we regard this as terrestrial
interference. Despite the RFI, Figure \ref{fig:iv_01417} shows that both
$I$ and $V$ are well-detected. In fitting Gaussians we are conservative
because we realize that the RFI may have contaminated the line shape. In
particular, the 0.33 MHz intervals happen to fall close to the two peaks in
Stokes $I$.

Table \ref{tab:01417} lists the parameters of the Gaussian components that
best fit the Stokes $I$ spectrum.  Column (1) lists the zero-based
component number.  Column (2) lists the peak flux density of each component
in mJy and the corresponding uncertainty.  Column (3) lists the central
heliocentric frequency of each component in MHz and the corresponding
uncertainty.  Column (4) lists the FWHM of each component in MHz and the
corresponding uncertainty.  Column (5) lists the optical heliocentric
velocity corresponding to the central frequency of each component.  Column
(6) lists the derived line-of-sight magnetic field in mG for each component
and the corresponding uncertainty.

Appying our guidelines from \S~\ref{sec:gauss} yielded the 5 Gaussian
components shown in Figure \ref{fig:iv_01417} and listed in Table
\ref{tab:01417}.  There are 2 distinct peaks in the OHM emission.  The peak
nearest 1622.8 MHz is asymmetric in such a way that 2 narrow components
needed to be added near this peak in order to minimize the residuals.  The
overall profile has a core-halo structure, with the 4 narrow components
lying on top of a broader component.

The Stokes $V$ spectrum ($S_{\rm rms} = 0.97$ mJy) in Figure
\ref{fig:iv_01417} shows prominent features that are fitted reasonably well
by the 5 Gaussian components, with Zeeman splitting yielding significant
fields in three Gaussians: Gaussian 0 has $B_{\parallel}=2.9 \pm 0.2$
mG, and Gaussians 3 and 4 have fields of $-2.7 \pm 0.1$ and $-3.6 \pm
0.3$ mG, respectively. Thus the field reverses from one peak to the
other.  \citet{pihlstromcbdp01} present 13 spectra from the 1667 MHz OHM
emission of III Zw 35 that they mapped using the EVN.  These data show
clearly that the 8215 and 8240 km s$^{-1}$ components (Gaussians 3 and 4)
arise from the southern peak, while the 8312 km s$^{-1}$ component
(Gaussian 0) is associated with the northern peak.  This provides clear
evidence that the reversal is arranged with the magnetic field pointing
away from us in the north and towards us in the south.

The panels on the right side of Figure \ref{fig:iv_01417} show results
relevant to the $B(\nu)$ fit. The top panel shows the composite
Gaussian-fitted (noise-free) Stokes $I$ spectrum. The center panel shows
the measured Stokes $V$ spectrum as a solid line; the dashed line
represents the Stokes $V$ spectrum that would be produced by a uniform
line-of-sight magnetic field of 1 mG --- this is obtained from equation
(\ref{eq:v}) by setting $B_{\parallel} = 1$ mG and using the derivative of
the composite Gaussian shown in the top panel.  The bottom panel shows the
$B(\nu)$ fit --- the derived $B_{\parallel}$ as a function of frequency ---
as described in \S~\ref{sec:circular_pol}. There is a clear systematic
pattern, with the field reversing sign from one peak to the other. The
estimated field strengths are also consistent with the Gaussian fits.

\citet{parracep05} and \citet{pihlstromcbdp01} present models of the OHM
emission in III Zw 35 as a clumpy, rotating starburst ring at an
inclination of 60$^{\circ}$, with an inner radius of 22 pc and a radial
thickness of 3 pc.  Both the Gaussian and $B(\nu)$ analyses suggest that an
azimuthal magnetic field is embedded within this starburst ring such that
the field points towards us at the southernmost tangent point and away from
us at the northernmost tangent point. \citet{parracep05} estimate that the
OHM clouds would be magnetically confined by a magnetic field of order
$\sim$ 10 mG.

\citet{killeensws96} used the ATCA to observe the OHM emission in III Zw
35.  Their sensivity of $S_{\rm rms} = 3.6$ mJy in Stokes $V$ was not
sufficient to detect the Zeeman splitting of the 1667 MHz line.

\subsubsection{IRAS F10038$-$3338 (IC 2545)}
\label{sec:v_10038}

As the residuals in Figure \ref{fig:iv_10038} show, the 1667 MHz OHM
emission from IC 2545 is fitted extremely well by 5 narrow Gaussian
components and one broad one.  Using our prescription from
\S~\ref{sec:gauss}, we see that there are 4 distinct peaks.  The peak
nearest 1613.1 MHz has an asymmetry that can be represented with a single
extra narrow component.  A broad component represents the evident core-halo
structure of the OHM emission profile.  It is unclear if the emission
feature at 1611.7 MHz corresponds to 1665 MHz emission or redshifted 1667
MHz emission.  The Stokes $I$ flux density and line profile have not
changed since the observations of \citet{killeensws96}.  Table
\ref{tab:10038} lists the Gaussian fit parameters shown in Figure
\ref{fig:iv_11506}.  The Stokes $V$ spectrum has an rms noise of $S_{\rm
  rms} = 0.7$ mJy.  There are three clear detections: Gaussian 1 probes a
field of $-1.8 \pm 0.3$ mG; Gaussian 2 is fitted by a field of $-11.3 \pm
1.2$ mG; and Gaussian 5 shows a reversal in sign with a field of $1.7 \pm
0.3$ mG.  Since no VLBI observations exist for this LIRG, nothing can be
said about the structure of the field reversal.

\subsubsection{IRAS F10173$+$0829}
\label{sec:v_10173}

As mentioned in \S~\ref{sec:rfi}, we used Hanning-smoothed spectra when
least-squares fitting this source because of RFI. We increased the derived
uncertainties in Table \ref{tab:10173} by the appropriate factor of
$\sqrt{8/3}$ \citep[Table A1]{killeensws96}.

\begin{figure*}[h!]
\begin{center}
  \includegraphics[width=3.5in] {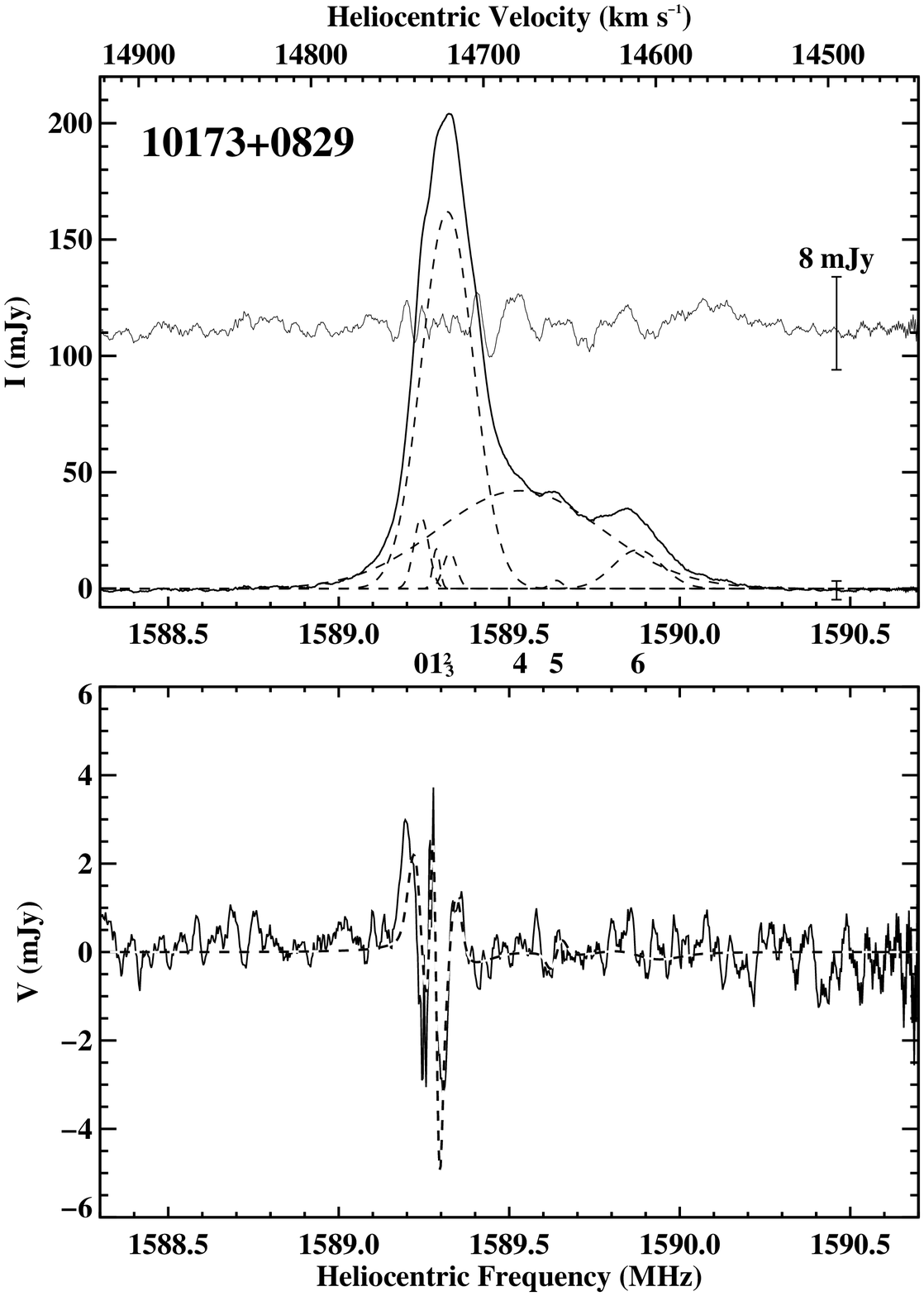}
  \includegraphics[width=3.5in] {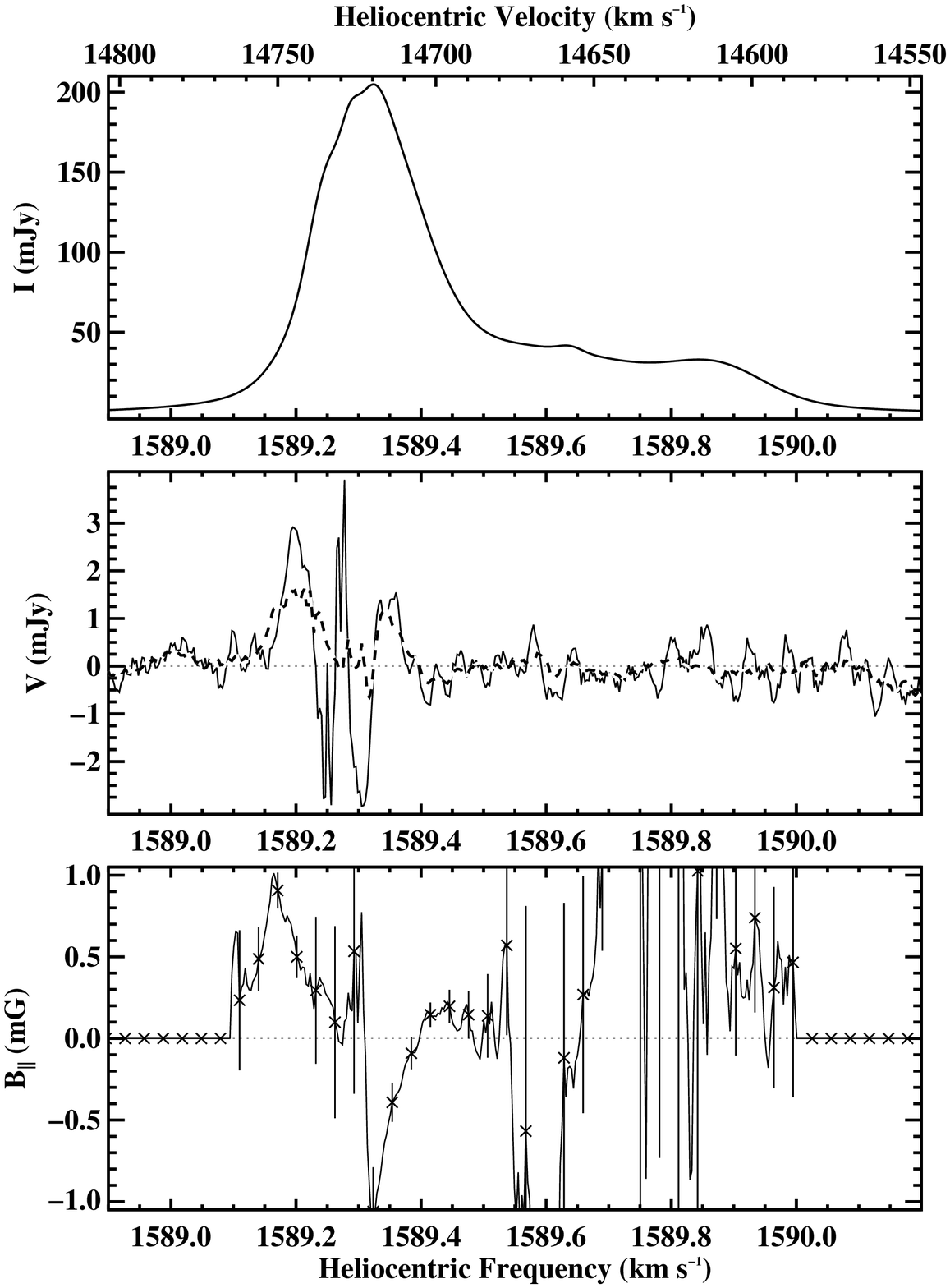}
\end{center}
\caption[Total intensity and circular polarization of IRAS
F10173$+$0829]{Total intensity and circular polarization results for IRAS
  F10173$+$0829. See caption for Figure \ref{fig:iv_01417}.  ({\it Top
    left}) Residuals are expanded by a factor of 5.  ({\it Middle right})
  Dashed line shows the hypothetical Stokes $V$ produced by a uniform
  $B_{\parallel} = 2$ mG using the derivative of the composite $I$ profile
  above.}
  \label{fig:iv_10173}
\end{figure*}

\begin{figure*}[h!]
\begin{center}
  \includegraphics[width=7in] {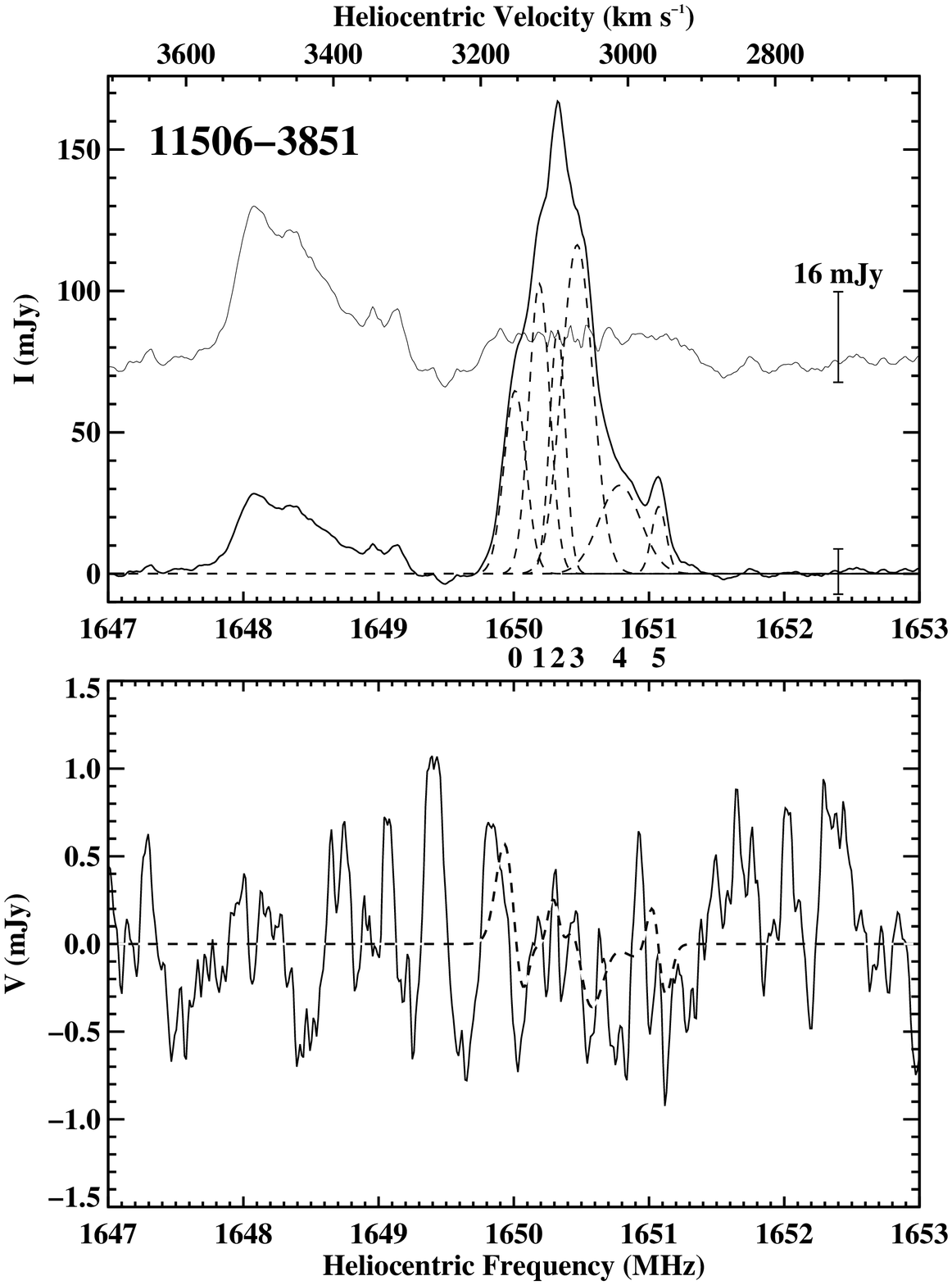}
\end{center}
\caption[Total intensity and circular polarization of IRAS
F11506$-$3851]{Total intensity and circular polarization results for IRAS
  F11506$-$3851. See caption for Figure \ref{fig:iv_01417}.  ({\it Top})
  Residuals are expanded by a factor of 2.}
  \label{fig:iv_11506}
\end{figure*}

We fit Stokes $I$ with 7 Gaussians as shown in Figure \ref{fig:iv_10173}
and listed in Table \ref{tab:10173}.  Using the selection guidelines of
\S~\ref{sec:gauss}, we required 4 narrow components to sufficiently fit the
extremely asymmetric peak near 1589.3 MHz.  The fit to Stokes $I$ yielded
reasonable residuals by including a single extremely broad component.  This
source represents a case where the profile structure is too complex to be
modelled by our straightforward fitting guidelines: we regard the derived
components as highly suspect regardless of the quality of the fit.

Detectable portions of the Stokes $V$ spectrum are restricted to the
strongly peaked $I$ line, where $V$ oscillates rapidly. Due to the
asymmetry in this profile, we found it impossible to obtain a set of
Gaussians that gives good fits to both $I$ and $V$.  Our final fit reflects
a compromise between large residuals and a larger number of Gaussians.
While it is clear that Zeeman splitting has been detected in this source,
we would need to add many more narrow, weak Gaussian components to our
Stokes $I$ fit in order to obtain statistically significant magnetic field
derivations from the fit to Stokes $V$ (which has an rms noise of $S_{\rm
  rms} = 0.8$ mJy).  Without any VLBI observations or any other physical
motivation for adding such components, the best we can do is present the
evidence for Zeeman splitting without estimating field strengths for
individual maser components.

Figure \ref{fig:iv_10173}, right panel, shows plots relevant to the $B(\nu)$
fit. The dashed line in the middle panel shows the Stokes $V$ spectrum that
would be expected for a uniform $B_{\parallel} = 2$ mG. The bottom panel
shows the $B(\nu)$ fit.  There is a clear systematic pattern, with the field
reversing sign from one side of the peak to the other. This reversal is not
revealed by the Gaussian fits because there are no narrow Gaussians on
either side of the peak and because the peak itself is represented by a
single Gaussian (component 2).

Since no spectral information is presented in the MERLIN maps of
\citet{yu04,yu05}, it is impossible to associate any of our Gaussian
components with OHM spots in 10173.

\subsubsection{IRAS F11506$-$3851}
\label{sec:v_11506}

The top panel of Figure \ref{fig:iv_11506} shows that the 1667 MHz Stokes
$I$ emission is fitted quite well by 6 narrow Gaussian components.  The
parameters for each component are listed in Table \ref{tab:11506}.
Gaussian components 0 and 3 have derived magnetic fields that look
significant when judged by their formal errors.  However, given the quality
of the Stokes $V$ spectrum ($S_{\rm rms} = 0.4$ mJy), we have no confidence in
either result. The 1665 MHz emission is clearly completely separated from
the 1667 MHz emission, yielding a hyperfine ratio $R_{H} = 4.9$.

\subsubsection{IRAS F12032$+$1707}
\label{sec:v_12032}

\begin{figure*}[h!]
\begin{center}
  \includegraphics[width=7in] {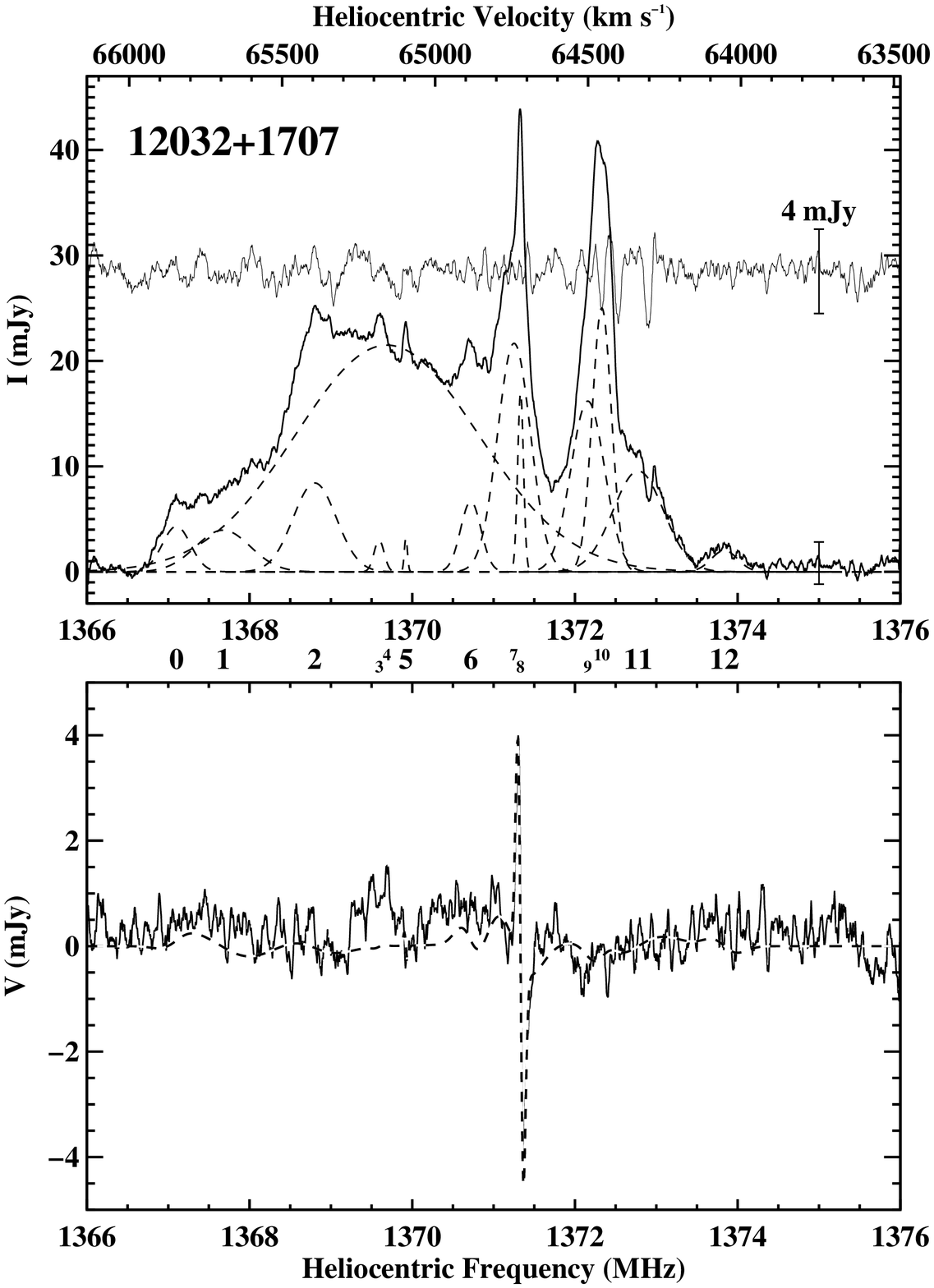}
\end{center}
\caption[Total intensity and circular polarization of IRAS
F12032$+$1707]{Total intensity and circular polarization results for IRAS
  F12032$+$1707. See caption for Figure \ref{fig:iv_01417}.  ({\it Top})
  Residuals are expanded by a factor of 2.}
  \label{fig:iv_12032}
\end{figure*}

\begin{figure*}[h!]
\begin{center}
  \includegraphics[width=7in] {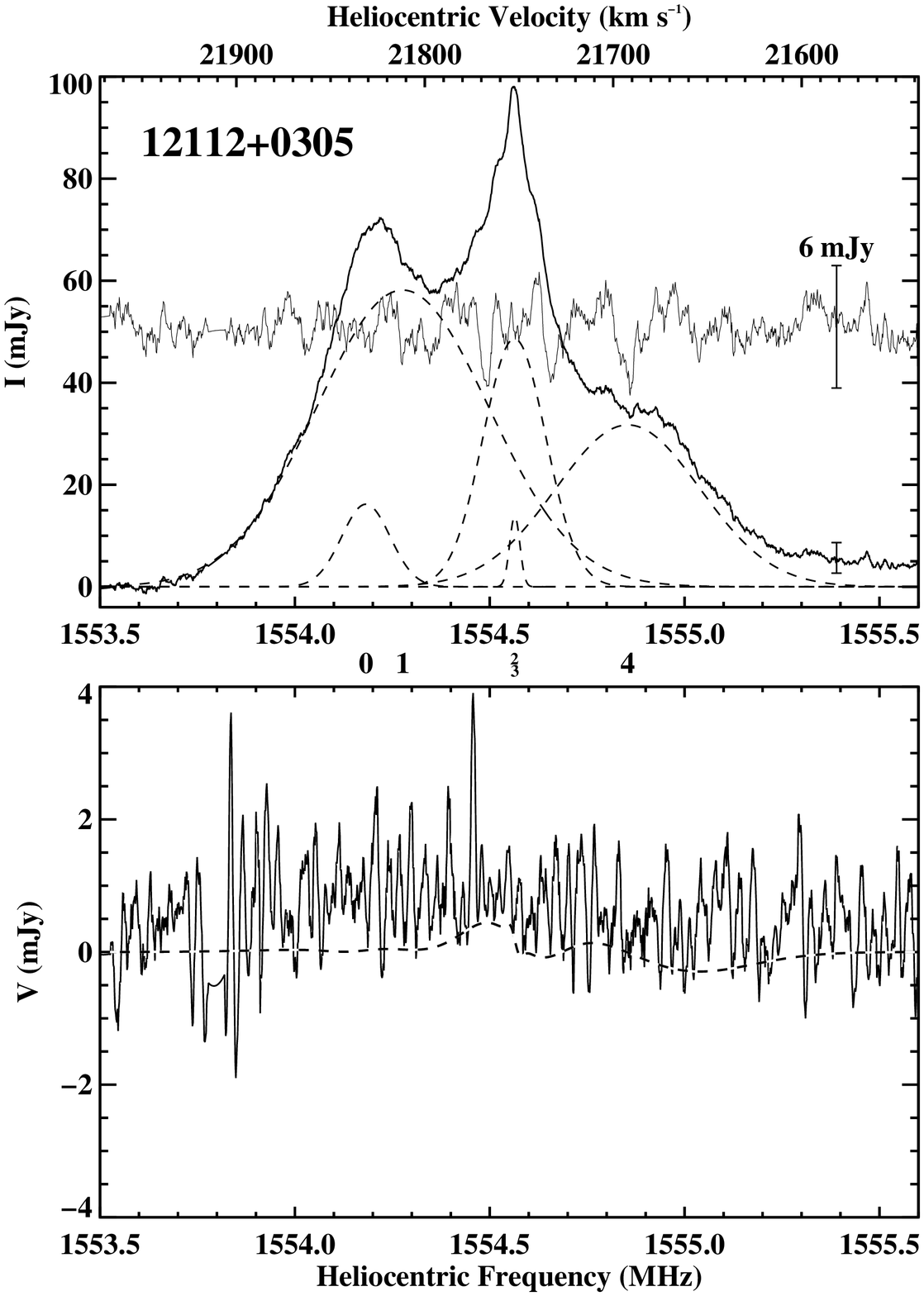}
\end{center}
\caption[Total intensity and circular polarization of IRAS
F12112$+$0305]{Total intensity and circular polarization results for IRAS
  F12112$+$0305. See caption for Figure \ref{fig:iv_01417}.  ({\it Top})
  Residuals are expanded by a factor of 4.}
  \label{fig:iv_12112}
\end{figure*}

\begin{figure*}[h!]
\begin{center}
  \includegraphics[width=7in] {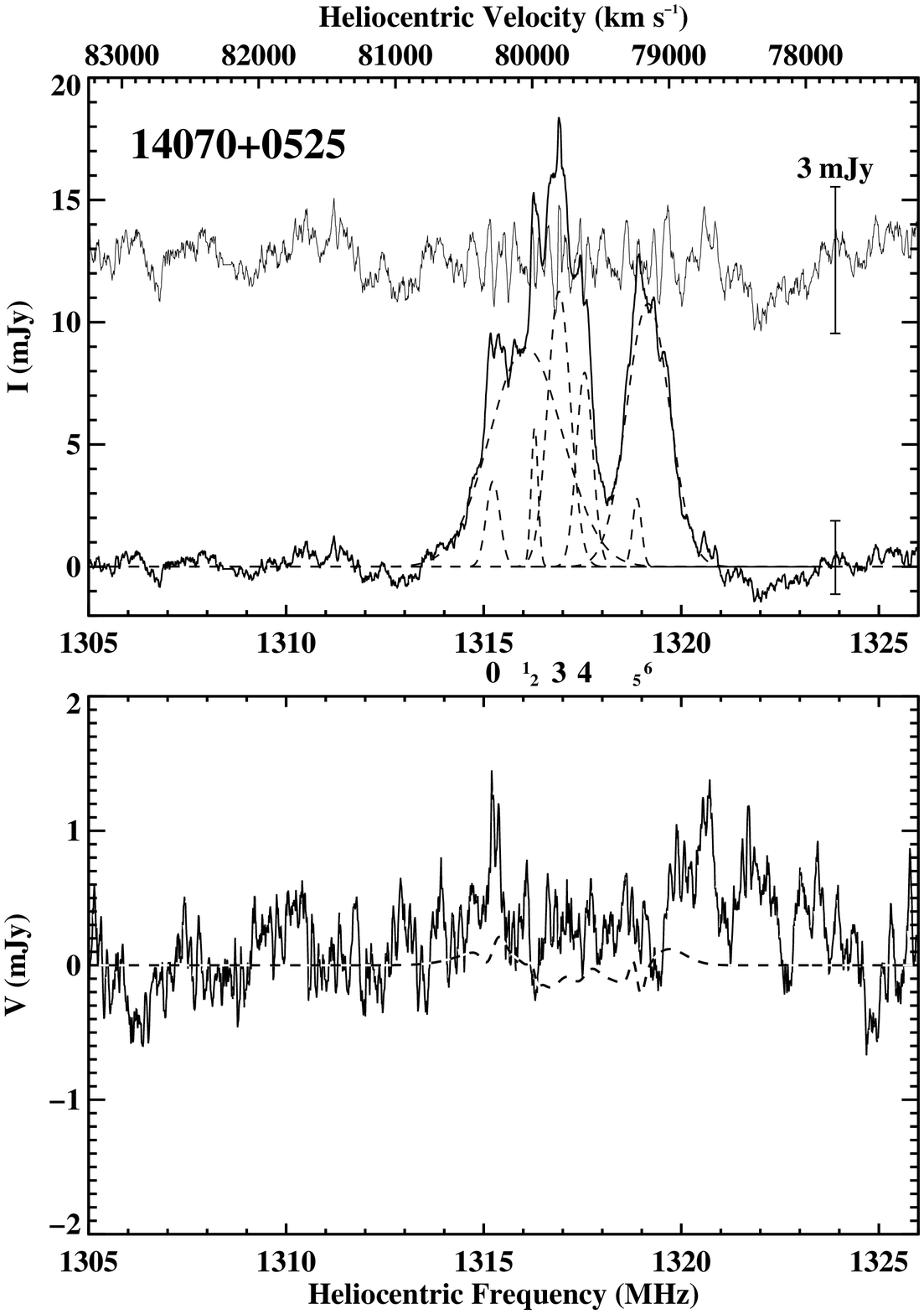}
\end{center}
\caption[Total intensity and circular polarization of IRAS
F14070$+$0525]{Total intensity and circular polarization results for IRAS
  F14070$+$0525. See caption for Figure \ref{fig:iv_01417}.  ({\it Top})
  Residuals are expanded by a factor of 2.}
  \label{fig:iv_14070}
\end{figure*}

The OHM emission from this gigamaser has an extremely wide extent; it is
impossible to distinguish the 1665 and 1667 MHz emission.  We fit its
Stokes $I$ spectrum with 13 Gaussians as shown in Figure \ref{fig:iv_12032}
with parameters as listed in Table \ref{tab:12032}.  Due to the complexity
of this profile, this is an extremely difficult case to apply our component
selection guidelines to.  We first identify 8 local maxima marked in Figure
\ref{fig:iv_12032} by components 0, 2, 3, 5, 6, 8, 10, and 12.  The peaks
near 8 and 10 display shoulders and are clearly blended with narrower
components: we added one additional component to each (components 7 and 9,
respectively).  We represent a broad shoulder, not quite intense enough to
produce a local maximum, near 1373 MHz by component 11. Finally, we add two
broad halo components, numbers 1 and 4, to minimize the residuals.  While
the prescription of \S~\ref{sec:gauss} allows us to select these 13
components somewhat straightforwardly --- even producing respectable
residuals --- visual inspection of Figure \ref{fig:iv_12032} suggests that
our model simplifies and glosses over the innate complexity of this OHM
profile.  Without VLBI observations, there is little that can be done to
improve upon our model.

Detectable signals in the $V$ spectrum ($S_{\rm rms} = 0.9$ mJy) are
restricted to the stronger of the two narrow peaks.  This narrow peak is
somewhat asymmetric and requires two Gaussians for a good fit. These two
components (numbers 7 and 8) have field strengths of $10.9 \pm 1.7$ and
$17.9 \pm 0.9$ mG, respectively.  Three other Gaussians, numbers 1, 6, and
11, have fields that are nearly $3\sigma$; however, visual examination of
the $V$ spectrum shows bumps and wiggles throughout at the $\approx$ 1 mJy
level, and we have no confidence in these purported fields.

Since the significant splittings occur where the Stokes $I$ profile has the
highest flux density, and since the 1667 MHz transition dominates in all
OHMs, we feel comfortable assuming that the emission is from the 1667 MHz
transition.  Of course, because of the blending of the 1665 and 1667 MHz
lines, there is an unresolvable ambiguity that could affect the derived
field strengths.

The Stokes $I$ line profile has changed since the source's discovery by
\citet{darlingg01}.  The flux density of the broad component and the narrow
component near 1372.3 MHz have remained the same.  However, the narrow
component near 1371.3 MHz, which used to be roughly 6 mJy (using our
classical definition of Stokes $I$) weaker than their 32.5 mJy peak at
1372.3 MHz, has flared and is now the strongest component with a flux
density of 44 mJy.  This time-variable component is the same one that
exhibits the largest splitting and therefore probes the strongest field; in
\S~\ref{sec:discussion} we compare this result with newly-observed strong
field detections in time-variable Galactic OH maser components.

\subsubsection{IRAS F12112$+$0305}
\label{sec:v_12112}

We were able to fit the Stokes $I$ OHM emission quite nicely with 5
Gaussians as shown in Figure \ref{fig:iv_12112}.  The fit parameters are
listed in Table \ref{tab:12112}.  There is no detectable signal in Stokes
$V$ ($S_{\rm rms} = 1.2$ mJy).  This is the first published spectrum of OHM
emission in 12112.

\subsubsection{IRAS F14070$+$0525}
\label{sec:v_14070}

Table \ref{tab:14070} lists the parameters for the 7 Gaussian components
used to fit the Stokes $I$ OHM emission in 14070. Since the 1665 and 1667
MHz lines are clearly blended in this source, we assume that all of the
components represent 1667 MHz emission.  As seen in Figure
\ref{fig:iv_14070}, this decomposition provides a decent fit but there is
no detectable signal in Stokes $V$ ($S_{\rm rms} = 0.5$ mJy). Gaussian
number 1 shows a nearly $3\sigma$ detection of magnetic field; however, the
associated feature in the Stokes $V$ spectrum appears to be no more
significant than the other features of mJy-strength intensity.  We have no
confidence in this near detection.

\subsubsection{IRAS F15327$+$2340 (Arp 220)}
\label{sec:v_15327}

\begin{figure*}[h!]
\begin{center}
  \includegraphics[width=7in] {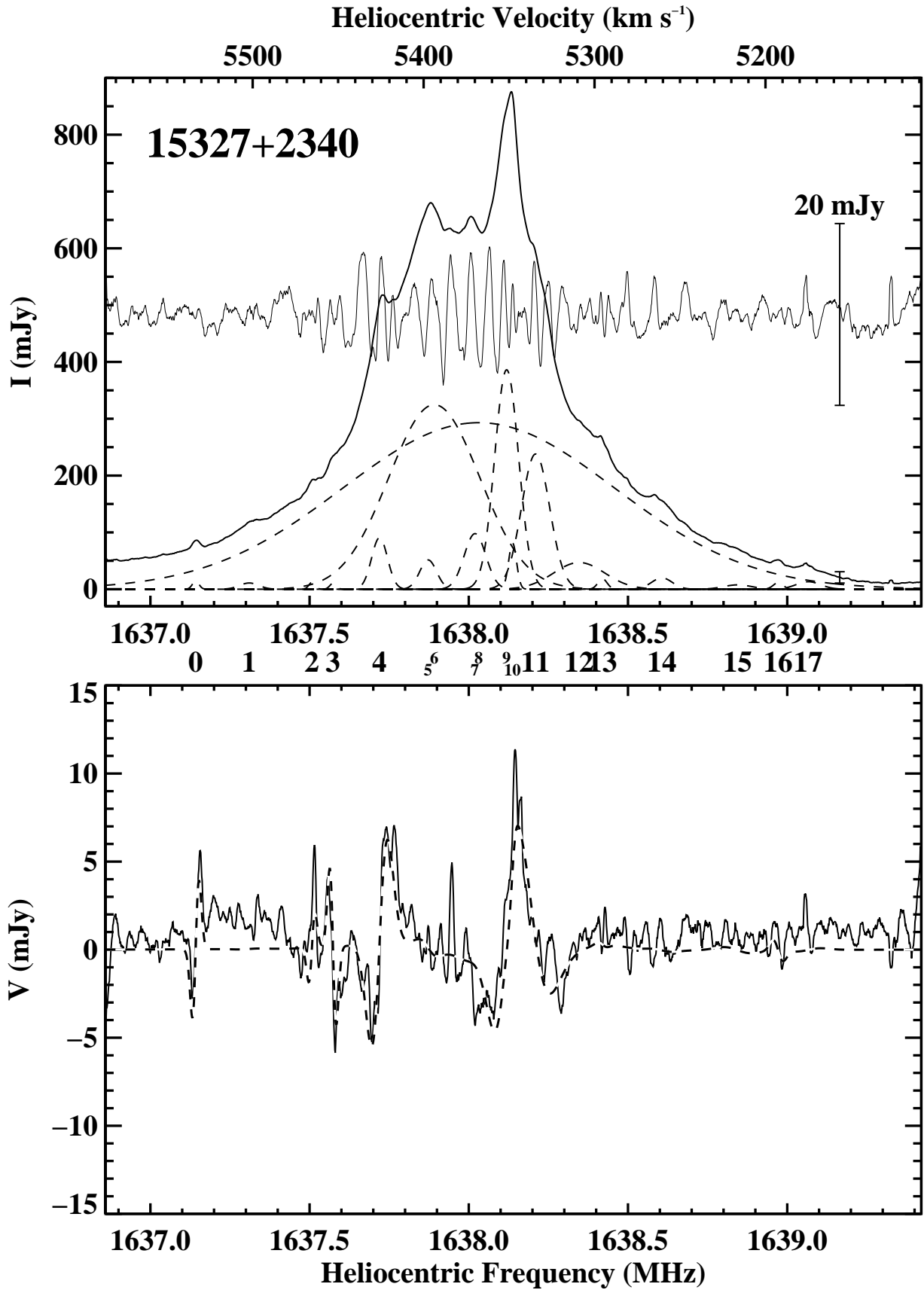}
\end{center}
\caption[Total intensity and circular polarization of IRAS F15327$+$2340
(Arp 220)]{Total intensity and circular polarization results for IRAS
  F15327$+$2340 (Arp 220). See caption for Figure \ref{fig:iv_01417}.
  ({\it Top}) Residuals are expanded by a factor of 16.}
  \label{fig:iv_15327}
\end{figure*}

The line profile of the Stokes $I$ OHM emission in Arp 220 is very complex.
Our fit required 18 Gaussian components, as seen in Figure
\ref{fig:iv_15327}, to obtain reasonable residuals.  An 18-component fit
might seem overwhelming, but the components are easily obtained using our
selection guidelines from \S~\ref{sec:gauss}.  There are 9 distinct peaks
(i.e., local maxima), represented in Figure \ref{fig:iv_15327} by Gaussian
components 0, 4, 5, 7, 9, 13, 14, 16, and 17. There are 6 narrow or fairly
narrow bumps or shoulders that are not intense enough to produce local
maxima, represented by Gaussian components 1, 2, 3, 11, 12, and 15.
Component 10 was needed to represent the asymmetry in the brightest peak
near 1638.15 MHz.  Finally, components 6 and 8 were needed to represent
core-halo structure in the overall profile.  This 18-component fit
reproduces all of the visually obvious narrow, weak bumps, as well as the
overall profile shape.  However, the residuals exhibit a different
signature in the line from that off the line, which means that our fit does
not represent the $I$ profile perfectly. We expended considerable effort
making sure that each of the 18 Gaussians listed in Table \ref{tab:15327}
is actually needed for the fit by inspecting the residuals for different
combinations of omitted Gaussians.

Six narrow Gaussians exhibit visually obvious signatures in Stokes $V$
(which has an rms noise of $S_{\rm rms} = 1.16$ mJy) and provide good fits
for Zeeman splitting. The absolute values of the derived field strengths
range from 0.7 to 4.7 mG, with 4 negative and 2 positive fields. Gaussians
2 and 3 have opposite field strengths.

Gaussian numbers 9 and 11 are strong (a few hundred mJy), have comparable
FWHMs of about 0.1 MHz, and are separated by about the FWHM.  This makes
them easily distinguishable.  They have opposite field directions as given
by the least-squares fit, but the reversal in sign is also visually
apparent. Zeeman splitting produces a Stokes $V$ pattern that looks like
the frequency derivative of the line, with amplitude and sign scaled by
$B_{\parallel}$. Thus, for a single Gaussian component, the $V$ pattern
looks like the letter ``S'' lying on its side, with an inevitable negative
and positive part; the integral over Stokes $V$ must be zero.  However, the
$V$ pattern for these Gaussians in Figure \ref{fig:iv_15327} doesn't look
like this; instead it is positive on both sides of the line and negative in
the middle. The only way to obtain positive $V$ on each side of a spectral
bump is for the field to have different signs on the two sides
\citep[cf.][Figure 3, the first radio detection of Zeeman splitting in Cas
A]{verschuur69}. The integral of Stokes $V$ must again be zero, with the
central negative portion balanced by the two positive ones on the sides.
The reversed field is not only a result of the fits, but is also visually
apparent.

We can compare our single-dish spectrum with the selected global VLBI
spectra presented by \citet{rovilosdlls03} and \citet{lonsdalelds98}.
There are a number of Gaussian components that appear to be directly
associated with the resolved OHM spots: component 11 at 5334 km s$^{-1}$
originates in a southwestern spot tracing a positive field; component 6 at
5393 km s$^{-1}$ originates in the southeast and traces a positive field;
component 4 at 5425 km s$^{-1}$ originates in the center of the northeast
ridge and traces a negative field; component 0 at 5533 km s$^{-1}$
originates in one of the southwestern spots tracing a negative field.
Three other features are more ambiguous: components 2 and 3 could be
associated with either the northwestern or northeastern OHM ridges, while
the brightest component, number 9 at 5351 km s$^{-1}$, appears to contain
emission from both the northern ridges as well as the southwestern maser
spots; these ambiguities prevent any possible field associations.  The
picture painted by the possible associations is for a field reversal from
positive to negative from the southern to the northern features of the
eastern OHM spots; there is no obvious reversal in the western region, but
it is possible given the associations above.


\subsection{Linear Polarization}
\label{sec:linear_pol}

For all observations, we used dual-polarized feeds with native linear
polarization. This means that the observed Stokes $U_{obs}$ and $V_{obs}$
come from cross-correlation products, which insulates them from system gain
fluctuations. However, Stokes $Q_{obs}$ comes from the difference between
the two native linear polarizations, so it is susceptible to time-variable,
unpredictable gain fluctuations. This leads to coupling between Stokes $I$
and Stokes $Q_{obs}$; in other words, a scaled replica of the $I$ profile
appears in the $Q_{obs}$ profile, with a random and unknown scaling factor,
so $Q_{obs}$ is unreliable.

Normally, when deriving linear polarization, one combines Stokes $Q_{obs}$
and $U_{obs}$ in the standard ways to obtain polarized intensity and
position angle for the astronomical source.  However, since $Q_{obs}$ is
unreliable for our measurements, we derived Stokes $Q_{src}$ and $U_{src}$
for the source from $U_{obs}$ alone by least-squares fitting its variation
with parallactic angle.  This is quite feasible at Arecibo because all
sources pass within $20^{\circ}$ of the zenith, so tracking for a
reasonably long time provides a wide spread in parallactic angle. This
makes the least-squares fit robust and provides good sensitivity and low
systematics.  For the source III Zw 35, which was plagued by serious
interference, we performed a minimum-absolute-residual-sum (MARS) fit.

As with the Stokes $I$ and $V$ spectra, the least-squares derived $Q_{src}$
and $U_{src}$ spectra are displayed after subtracting both the off-source
position and the continuum.  We then use these baseline-subtracted Stokes
spectra to derive spectra for polarized intensity and position angle. We do
this because, even for the position-switched spectra, the continuum linear
polarization is usually dominated by the diffuse Galactic synchrotron
background. Although this prevents us from reliably deriving linear
polarization for the (U)LIRG continuum radiation, the frequency-variable
polarization is reliable.

For two sources below, we least-squares fit for the Faraday rotation
measure $RM$. Performing this fit requires some care because the $RM$ is
derived from the position angle $\psi$, which in turn is obtained by
combining $Q_{src}$ and $U_{src}$, which combine nonlinearly through the
$\arctan$ function [$\psi = 0.5 \arctan(Q_{src}/U_{src})$]. The
channel-by-channel data are too noisy to produce a good-looking spectral
plot of $\psi$, so on our plots we boxcar smooth by an appropriate number
of points. One cannot linearly fit the unsmoothed values of $\psi$ to
frequency because the $\arctan$ function produces nonlinear noise in
$\psi$. To avoid this problem, we performed a nonlinear fit to the
unsmoothed $\arctan(Q_{src}/U_{src})$; this extra complication ensures that
the derived values and errors are unaffected by smoothing.

We were unable to analyze the linear polarization for the two sources we
observed using the GBT (IC 2545 and 11506) because of inadequate
parallactic angle coverage.  We report the linear polarization for the
Arecibo results here and discuss their interpretation in
\S~\ref{sec:discussion_faraday}.

We present three vertically stacked plots for each source below (Figures
\ref{fig:01417qandu}--\ref{fig:15327qandu}).  The top panel shows the
position-differenced, baseline-subtracted Stokes $I$ spectrum over 12.5
MHz, therefore including both the 1665 and 1667 MHz transitions for each
source.  The middle panel presents the linear polarization intensity and
the bottom panel displays the derived position angle $\psi$ as a function
of heliocentric frequency.

\subsubsection{IRAS F01417$+$1651 (III Zw 35)}
\label{sec:lin_01417}

\begin{figure*}[h!]
\begin{center}
  \includegraphics[width=3.5in] {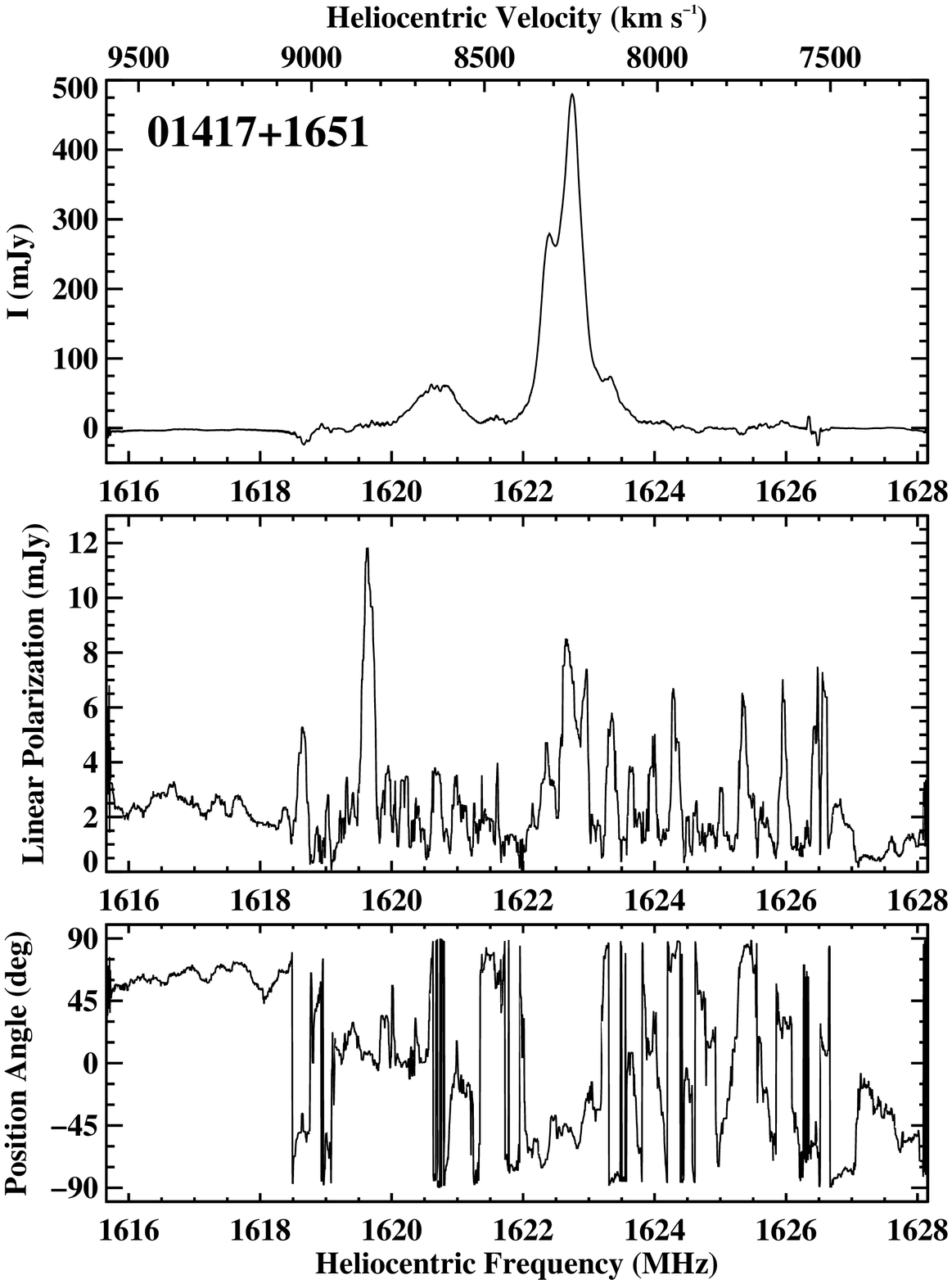}
  \includegraphics[width=3.5in] {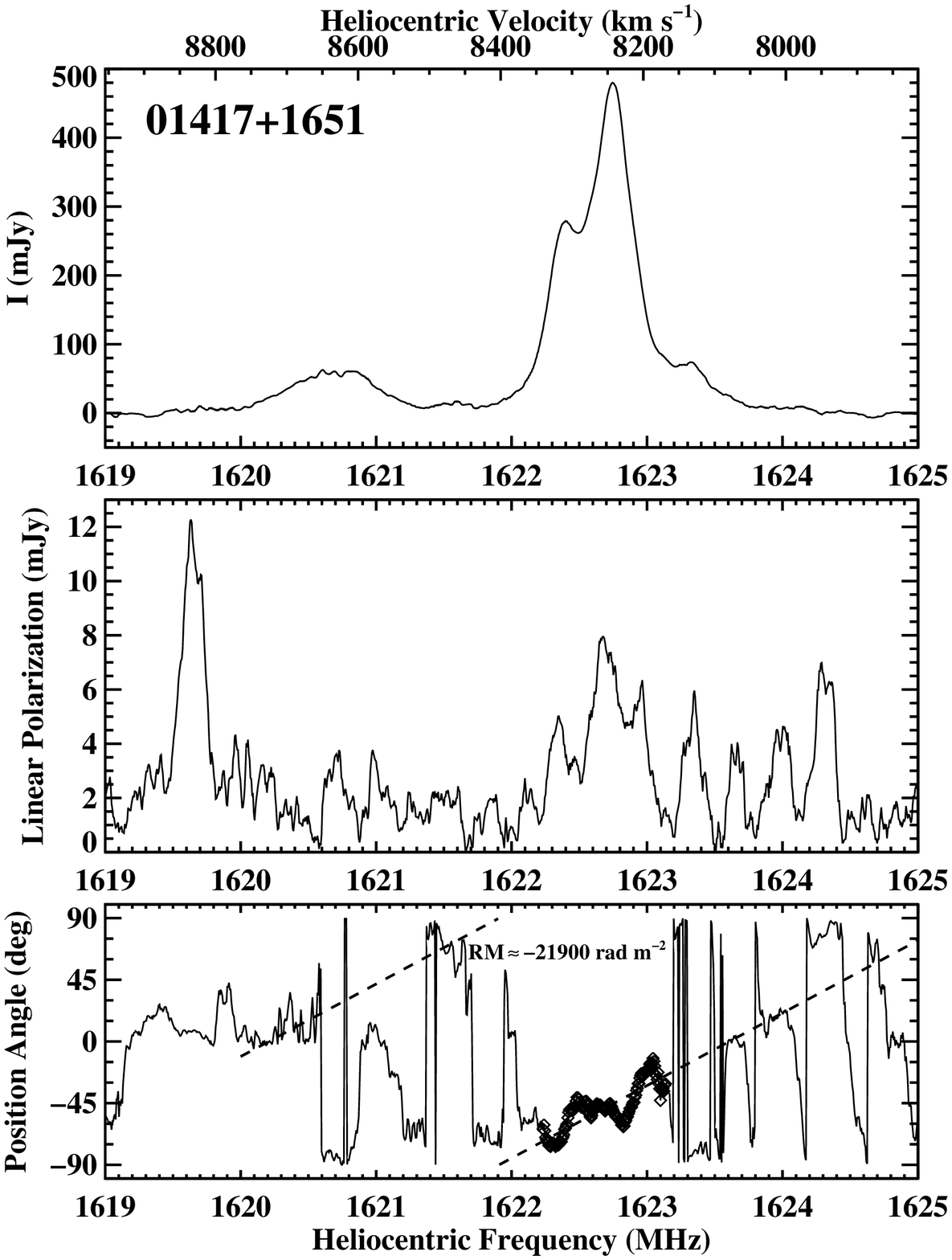}
\end{center}
\caption[Linear polarization of IRAS F01417$+$1651 (III Zw 35)]{Linear
  polarization results for IRAS F01417$+$1651 (III Zw 35).  ({\it Left})
  Top panel shows Stokes $I$, middle panel shows the linearly polarized
  intensity, and bottom panel shows the position angle over the entire 12.5
  MHz bandwidth. ({\it Right}) Same as left panels, but with the frequency
  range narrowed to 6 MHz; the bottom right plot also shows the fitted
  Faraday rotation as a dashed line whose slope was determined by fitting
  to the points marked as diamonds.  All spectra are plotted as a function
  of heliocentric frequency ({\it bottom axis}). The top panels show the
  optical heliocentric velocity ({\it top axis}). All spectra are smoothed
  by a boxcar of 23 channels.}
  \label{fig:01417qandu}
\end{figure*}

\begin{figure*}[h!]
\begin{center}
  \includegraphics[width=7in] {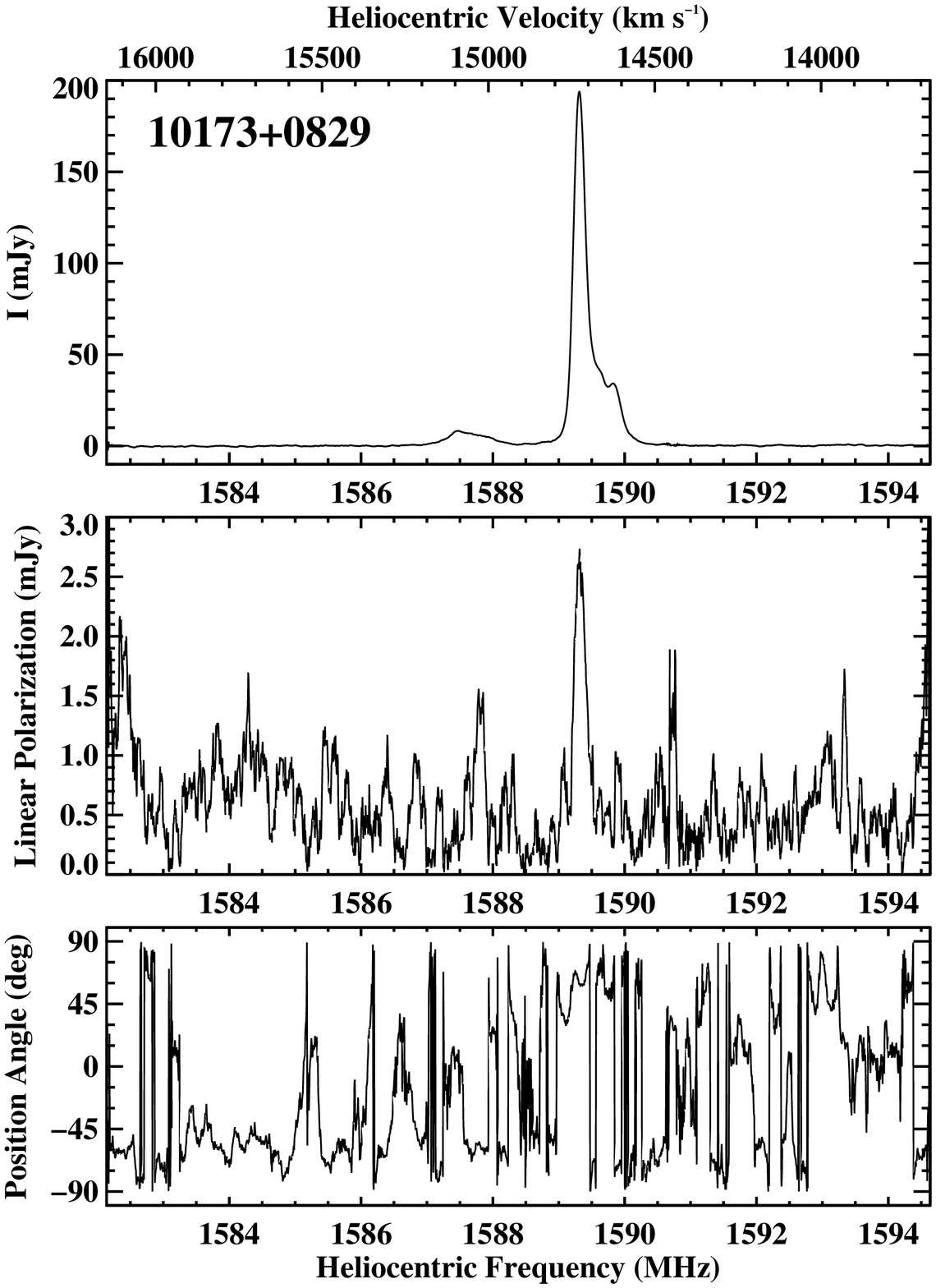}
\end{center}
\caption[Linear polarization of IRAS F10173$+$0829]{Linear polarization
  results for IRAS F10173$+$0829. See caption for Figure
  \ref{fig:01417qandu}. All spectra are smoothed by a boxcar of 17
  channels.}
  \label{fig:10173qandu}
\end{figure*}

\begin{figure*}[h!]
\begin{center}
  \includegraphics[width=7in] {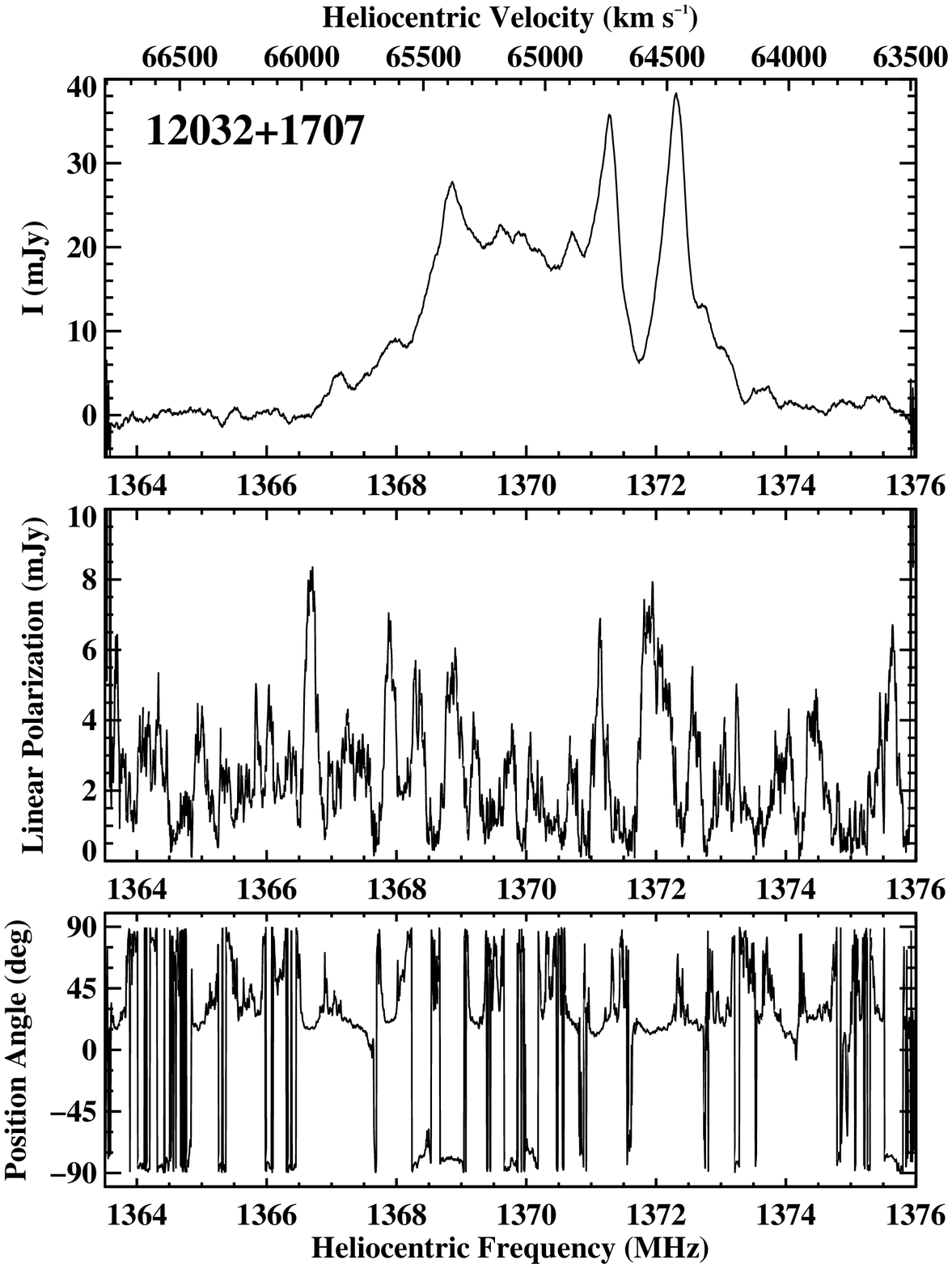}
\end{center}
\caption[Linear polarization of IRAS F12032$+$1707]{Linear polarization
  results for IRAS F12032$+$1707. See caption for Figure
  \ref{fig:01417qandu}. All spectra are smoothed by a boxcar of 31
  channels.}
\label{fig:12032qandu}
\end{figure*}

\begin{figure*}[h!]
\begin{center}
  \includegraphics[width=7in] {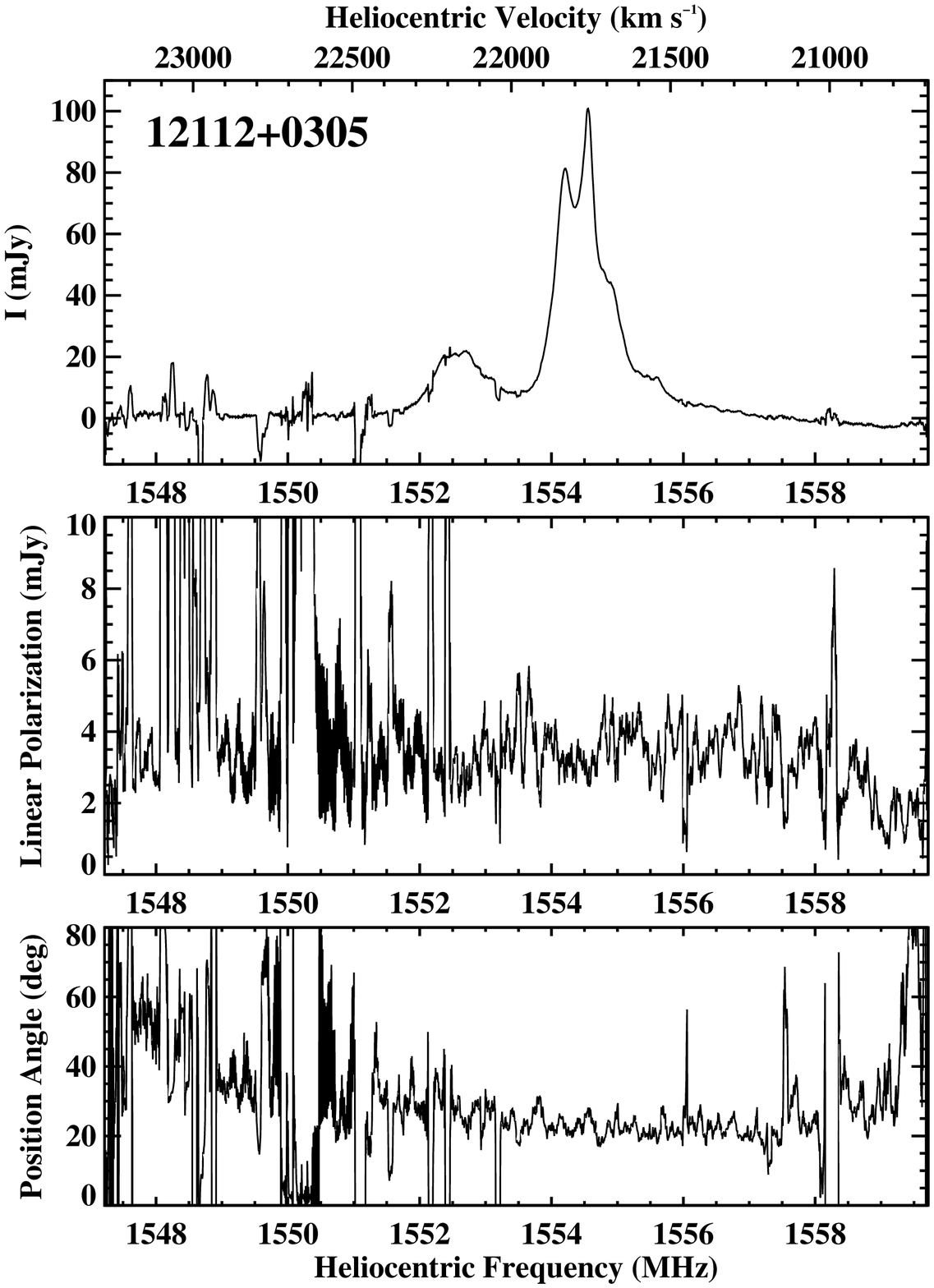}
\end{center}
\caption[Linear polarization of IRAS F12112$+$0305]{Linear polarization
  results for IRAS F12112$+$0305. See caption for Figure
  \ref{fig:01417qandu}. All spectra are smoothed by a boxcar of 11
  channels.}
\label{fig:12112qandu}
\end{figure*}

\begin{figure*}[h!]
\begin{center}
\includegraphics[width=7in] {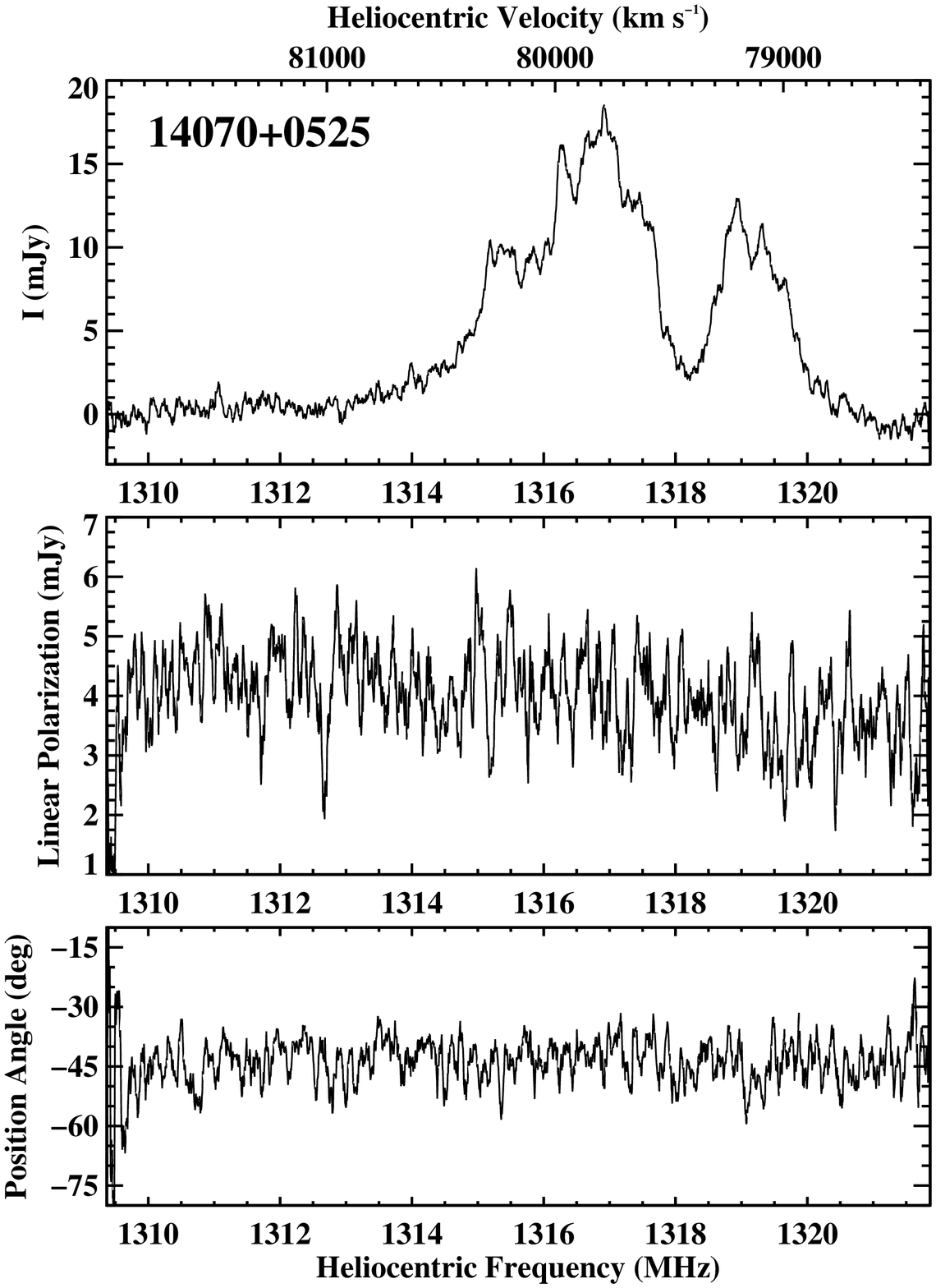}
\end{center}
\caption[Linear polarization of IRAS F14070$+$0525]{Linear polarization
  results for IRAS F14070$+$0525. See caption for Figure
  \ref{fig:01417qandu}. All spectra are smoothed by a boxcar of 11
  channels.}
  \label{fig:14070qandu}
\end{figure*}

The linear polarization results for III Zw 35 are presented in Figure
\ref{fig:01417qandu}. The top panels exhibit the position-differenced,
baseline-subtracted Stokes $I$ spectrum.  The 1665 MHz transition is
clearly visible and the hyperfine line ratio is $R_{H} = 6.0$.  The middle
panels clearly display that the spectrum of linearly polarized intensity is
extremely spikey.  Although higher S/N would help, the spikes might be real
and possibly correspond to individual masers that are too weak to be seen
clearly in the Stokes $I$ spectrum.  The polarized intensity shows a
seemingly real peak centered near 1622.8 MHz, which is also the center of
the Stokes $I$ peak. The polarized intensity is about 5 mJy and the Stokes
$I$ peak is roughly 500 mJy, so the fractional polarization $\approx 1\%$.
If the other spikes are real, then their fractional polarizations are much
higher.

The bottom panels display the position angle $\psi$. Position angles
exhibit less scatter than intensities, and the angle looks well-defined for
the 1622.8 MHz peak. Also, it seems to show a gradual change across the
line, which is about 1 MHz wide. The dashed line displays the result of a
least-squares fit to the frequency dependence of the position angle, using
only those points that are marked as diamonds: $RM = -21900 \pm 3700$ rad
m$^{-2}$. The extrapolated dashed line goes through the clusters of points
associated with spikes centered near 1623.7 and 1624.0, and moreover, even
the slope of the line matches the data for these spikes. The slope also
seems to match the 1624.4 MHz polarized-intensity spike, but the data are
offset by about $60^{\circ}$. We speculate that: (1) these three
polarized-intensity spikes come from individual OHMs that are too weak to
see in the top panels of Figure \ref{fig:01417qandu}; (2) they all suffer
the same Faraday rotation of $\simeq -21900$ rad m$^{{-2}}$ as the central
peak; and (3) the intrinsic position angle for the 1624.4 MHz maser differs
from the other two by about $60^{\circ}$.

\subsubsection{IRAS F10173$+$0829}

Figure \ref{fig:10173qandu} displays the linear polarization results for
10173. The polarized intensity shows a low-S/N spike that is centered on
the Stokes $I$ line: the polarization fraction is about $1\%$ and the
position angle about $60^{\circ}$. The spike is too narrow to fit for
Faraday rotation.  The hyperfine line ratio for 10173 is $R_{H}=10.7$.

\subsubsection{IRAS F12032$+$1707}

The linear polarization results for 12032 are shown in Figure
\ref{fig:12032qandu}. The polarized intensity shows multiple spikes that
might be real. The most significant is centered at $\simeq 1372$ MHz, with
a peak flux density of $\simeq 5$ mJy, and has $\psi \simeq 15^{\circ}$;
near this frequency, $I$ varies from $\simeq 10$ to $\simeq 40$ mJy, so if
this peak is real then the fractional polarization is huge, $\approx 50\%$
to $\approx 10\%$ --- unheard of for OH masers of any stripe.

\subsubsection{IRAS F12112$+$0305}

Figure \ref{fig:12112qandu} shows the linear polarization results for
12112. The lower frequencies are plagued by RFI, which remarkably
disappears at the low-frequency boundary of the 1667 MHz line (centered at
1554.5 MHz).  According to the National Telecommunications and Information
Administration Manual of Regulations and Procedures for Federal Radio
Frequency Management, this RFI is likely attributable to space-to-Earth
aeronautical mobile satellite communications operated by Inmarsat. There is
no trace of any detectable linear polarization for this source. This is the
first detection of the 1665 MHz transition for 12112; the hyperfine line
ratio is $R_{H} = 4.0$.

\subsubsection{IRAS F14070$+$0525}

Figure \ref{fig:14070qandu} displays the linear polarization results for
14070.  The linear polarization intensity is approximately $4$ mJy across
the entire 12.5 MHz bandwidth with an estimated position angle of
$-48^{\circ}$.

\subsubsection{IRAS F15327$+$2340 (Arp 220)}

\begin{figure*}[h!]
\begin{center}
  \includegraphics[width=3.5in] {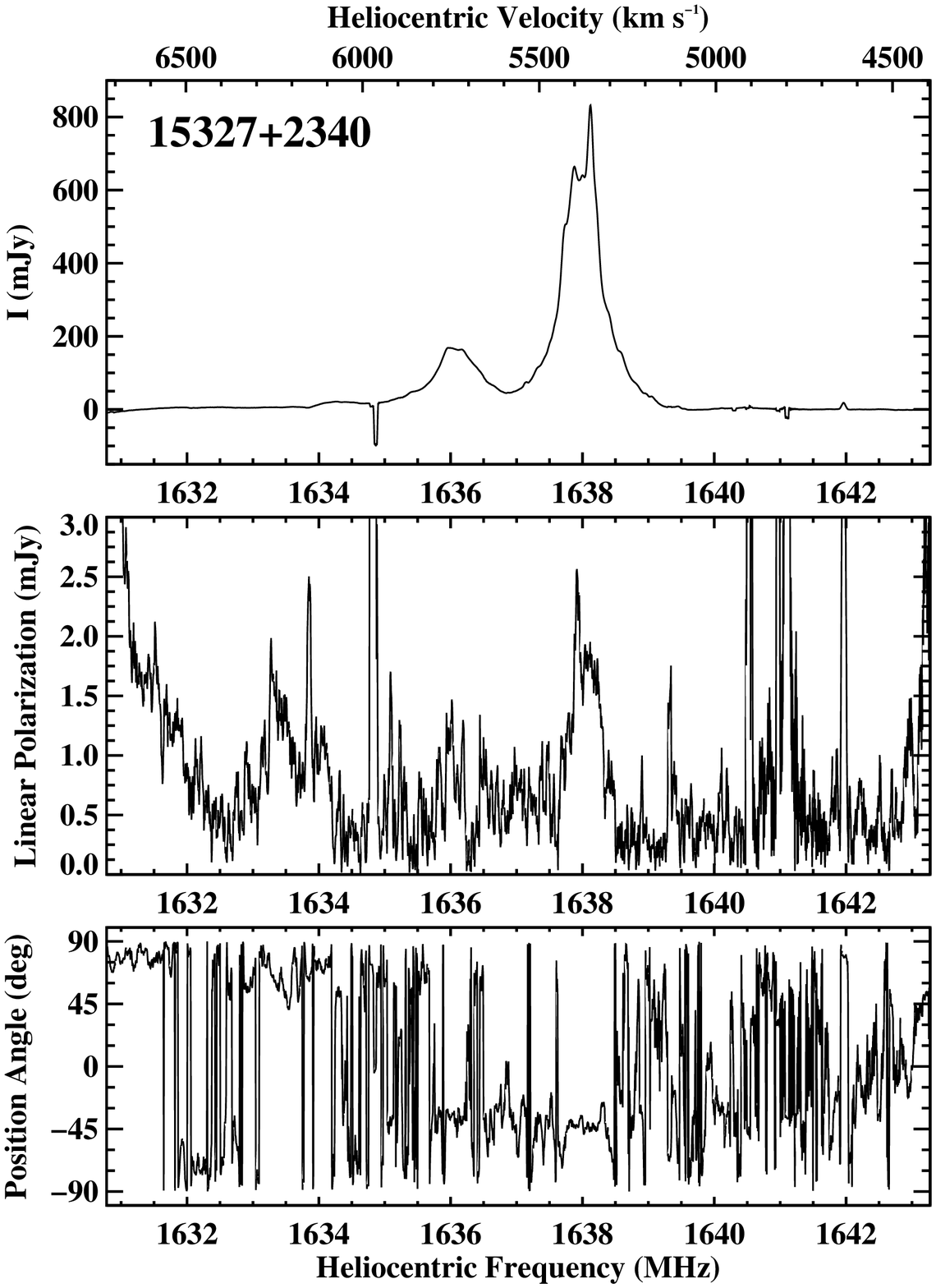}
  \includegraphics[width=3.5in] {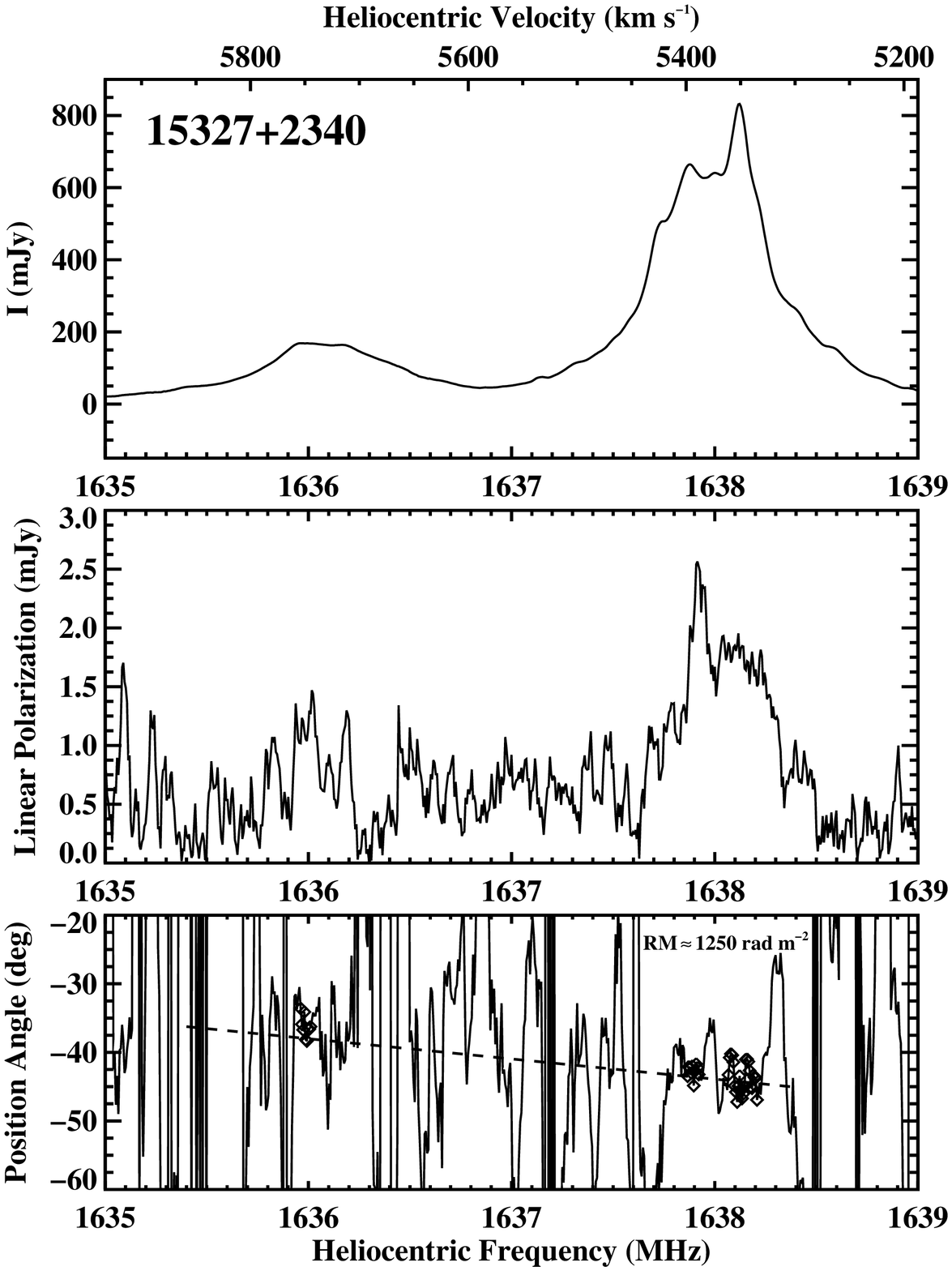}
\end{center}
\caption[Linear polarization of IRAS F15327$+$2340 (Arp 220)]{Linear
  polarization results for IRAS F15327$+$2340 (Arp 220).  See caption for
  Figure \ref{fig:01417qandu}. ({\it Right}) The frequency range has been
  narrowed to 4 MHz. All spectra are smoothed by a boxcar of 9 channels.}
\label{fig:15327qandu}
\end{figure*}

The top panels of Figure \ref{fig:15327qandu} show the Stokes $I$ profile
for Arp 220 including both the 1665 and 1667 MHz transitions.  The
hyperfine line ratio is $R_{H}=3.5$. The middle panels show the linear
polarization intensity, which has a well-defined peak centered at 1638 MHz
and peaks at about 2 mJy. This is only $\approx 0.3\%$ of the total
intensity at this frequency. This is a very small fractional polarization
but is very well-detected.

The bottom panels show that the position angle of linear polarization is
well-defined in two regions of low noise, one centered near 1638 MHz and
the other near 1636 MHz.  The former region corresponds to the 1667 MHz
line and the latter is aligned with the 1665 MHz transition. This 1665 MHz
line is unconvincingly visible in the polarized intensity spectrum, but the
low noise in its position angle spectrum is unmistakable.

We fit the frequency variation of $\psi$ to obtain the Faraday rotation
measure $RM$ using those points marked as diamonds in the bottom right
panel of Figure \ref{fig:15327qandu}.  For the 1638 MHz component alone, we
obtain $RM = 5230 \pm 7930$ rad m$^{-2}$. For the combination of the 1636
and 1638 MHz components, we obtain $RM = 1250 \pm 1040$ rad m$^{-2}$. These
errors are considerable and make the formal result only marginally
significant.  The dashed line in the bottom-right panel displays the result
of the fit for both components together; visual inspection shows that not
only is it an acceptable fit for both components together, but it is also
acceptable for the 1638 MHz component alone. It is not unreasonable to
conclude that the OHM radiation from both OH lines suffers a common Faraday
rotation of $RM \approx 1250$ rad m$^{-2}$; this is $\sim$ 20 times smaller
than the value derived for III Zw 35.


\section{Discussion}
\label{sec:discussion}

\subsection{OH Maser Zeeman Pairs in the Milky Way}
\label{sec:discussion_mw}

While our results are the very first {\it in situ} Zeeman detections in
external galaxies, OH masers in the MW have been used as Zeeman
magnetometers for well over a decade. In contrast to OHMs, Galactic OH
maser emission lines are so narrow ($\sim 0.5\ {\rm km\ s^{-1}}$) that
fields of $\approx 1$ mG are sufficient to completely split the left and
right circular $\sigma$ components into pairs.  More than 100 of these
Zeeman pairs have been compiled by \citet{fishram03} and \citet{reids90}
with a distribution whose mean is consistent with 0 mG and whose standard
deviation is 3.31$\pm$0.09 mG.  Typical densities in OH maser regions are
$n \sim 10^{6}-10^{7}\ {\rm cm}^{-3}$; for a field strength of $\sim 10\
\mu$G in gas at $\sim 1-100$ cm$^{-3}$, the fields probed by Galactic OH
masers are consistent with the enhancement of $|B| \propto n^{1/2}$
\citep{fishram03}.  The linear polarization of the $\sigma$ components is
often measured in addition to the $\pi$ component, but the $\pi$
components, which are in theory 100\% linearly polarized, are rarely
measured to be purely so.

Unlike in the OH masers in our Galaxy, the flux density of the 1667 MHz
transition in all OHMs is larger than that of the 1665 MHz transition and,
until now, no polarization has been detected \citep{lo05}. There is no
definitive explanation for the dominance of the 1667 MHz transition, but
recent work suggests that this probably arises because the extragalactic
lines are wider than the Galactic maser lines (P.\ Goldreich, private
communication; M.\ Elitzur, private communication; \citealp{lockette08}).

Our detections yield a median line-of-sight magnetic field strength of
$\simeq$ 3 mG in OHMs in (U)LIRGs, which is comparable to the field
strengths measured in OH masing regions in the MW.  This strongly suggests
that the {\it local} process of massive star formation occurs under similar
conditions in (U)LIRGs, galaxies with vastly different large-scale
environments than our own.

The magnetic field strengths we find in the OHMs in (U)LIRGs ($\sim 3$ mG)
are comparable to the volume-averaged fields of $\gtrsim 1$ mG inferred
from synchrotron observations.  These results imply that mG magnetic fields
likely pervade most phases of the interstellar medium in (U)LIRGs.  It is
unclear, however, how to physically relate the two different magnetic field
strengths in more detail given the possibility that each may probe rather
different phases of the ISM.  Some models of OHMs invoke radiative pumping
in molecular clouds with gas densities $\sim 10^{3.5}-10^4$ cm$^{-3}$
\citep[e.g.,][]{randellfjyg95}.  This is similar to the {\it mean} gas
density in the central $\sim 100$ pc in (U)LIRGs, in which case our
observations likely probe the mean ISM magnetic field (whether the
synchroton radiation also arises from gas at this density is unclear;
upcoming GLAST observations of neutral pion decay may help assess this; see
\citealp{thompsonqw07}).  It is also possible, however, that the OHMs arise
in somewhat denser gas ($n \sim 10^6-10^7$ cm$^{-3}$; e.g.,
\citealp{lonsdalelds98}), as appears to be true in the MW
\citep[e.g.,][]{fishram03}.  In this case, the magnetic field probed by
OHMs is likely stronger than that in the bulk of the ISM.  If we assume the
$B \propto n^{1/2}$ scaling often assumed in the MW
\citep{mouschovias76,fishram03}, the field strengths in the masing regions
in (U)LIRGs are probably within a factor of $\sim 3$ of the mean ISM field
(rather than a factor of several 100 in the MW), given the large mean gas
densities in (U)LIRGs.  This is still reasonably consistent with the mean
field strength of $\gtrsim 1$ mG inferred from synchrotron observations.
Without a better understanding of the physical conditions in the masing
regions, however, it is difficult to provide a more quantitative connection
between our inferred field strengths and either the mean ISM field or the
magnetic field probed by synchrotron emission.  Ultimately doing so is
important because it will allow stringent constraints to be placed on the
dynamical importance of magnetic fields across a wide range of physical
conditions in (U)LIRGs.

\citet{fishram03}, with their comprehensive survey of Galactic OH masers
and the accompanying statistical discussion, strongly support several
previous suggestions that the field {\it direction} in OH masers usually
mirrors that of the large-scale field in the vicinity of the masers.
MERLIN observations of OH masers in Cep A by \citet{bartkiewiczscr05}
also present Zeeman detections corroborating the field's alignment with the
ambient ISM field direction. Thus, measuring the direction of the field in
an OH maser reveals the field direction not only {\it in} the maser, but
also {\it outside and in the vicinity of} the OH maser. For the MW, this
aids us to infer the large-scale magnetic field morphology.  To directly
compare the star formation processes in the MW and (U)LIRGs, it will be
necessary to increase the sample of magnetic field strengths in (U)LIRGs and
to directly map the Zeeman splitting of individual OHM spots using VLBI in
order to probe whether reversals occur at smaller angular scales.

\subsection{Strong Fields and Time Variability}
\label{sec:discussion_time_variability}

\citet{slyshm06} and \citet{fishr07} both observed fields of 40 mG using
Zeeman observations of OH maser spots in W75N; these are the highest field
strengths measured in Galactic OH masers and are an order of magnitude
larger than the typical OH maser field.  These OH maser spots also happen
to have been flaring based on multi-epoch VLBA observations; perhaps time
variability in OH masers is correlated with strong magnetic fields.
Interestingly, our strongest detection, $B_{\parallel} \sim 18$ mG in the
gigamaser 12032, occurs in an OHM component that has increased in flux
density by a factor of two since its previous published observation
\citep{darlingg01}.

These results strongly support the development of an observational program
to monitor both the time variability of the Stokes $I$ flux density and
magnetic field strength in OHMs as well as the necessity of observing the
circular polarization of time-variable Galactic masers in hopes of
detecting strong magnetic fields.

\subsection{Linear Polarization and Faraday Rotation}
\label{sec:discussion_faraday}

Our measured rotation measures of $RM \simeq 21900$ rad m$^{-2}$ for III Zw
35 and $RM \simeq 1250$ rad m$^{-2}$ for Arp 220 are large by most
standards, but are not unreasonable for (U)LIRGs.  As mentioned in
\S~\ref{sec:intro}, the magnetic field strength throughout the ULIRG ISM
should be $\gtrsim 1$ mG from synchrotron observations.  Electron densities
are estimated to be $\sim 1-10$ cm$^{-3}$ in the hot ionized plasma, both
from observations of X-ray emission \citep[e.g.,][]{grimeshsp05} and from
theoretical models of supernova-driven galactic winds
\citep[e.g.,][]{chevalierc85}. Over a path length $\sim 100$ pc in the
central portions of ULIRGs, $n_e \sim 1$ cm$^{-3}$ and $B \sim 1$ mG imply
$\langle n_{e}B_{\parallel} L\rangle \sim 0.1$ G cm$^{-3}$ pc, or $RM \sim
80000$ radians m$^{-2}$. This is a factor of 4--60 larger than our measured
values.

It is reasonable for this simple estimate to overestimate the measured
$RM$.  This is because the $RM$ depends only on the line-of-sight field
component. The probability density function for the line-of-sight component
of a randomly oriented magnetic field is flat between zero and the
perfectly-oriented case; thus, for a set of sources with randomly-oriented
fields, the observed line-of-sight field component is reduced by a factor
of two, and 1/4 of the sources have the observed component less than 1/4 the
perfectly-aligned value.  In addition, and probably more importantly, the
observed Faraday rotation responds only to the systematic line-of-sight
field component, while the synchrotron radiation and the total magnetic
energy depend on the total field, systematic plus random. Our estimate of
$RM \sim 80000$ is for the total field, not the systematic field, because
the latter is much harder to predict.

The measured RM might also be reduced by finite source-size effects and/or
propagation through an inhomogenous medium \citep{burn66}.  First, suppose
that the magnetic field is everywhere uniform but that the Faraday rotation
is produced in the same region where the maser radiation is produced, and
that this region is extended along the line of sight. In this case,
different line-of-sight depths of the maser are rotated by different
amounts. This washes out the linear polarization and can reduce the
apparent Faraday rotation.  In the other extreme, think of the field as
primarily random except for a small uniform component.  Maser radiation
observed at a given frequency might come from more than one maser located
at different positions on the sky or at different distances into the
source. In the former case, the $RM$ might change with position on the sky;
in the latter, it might change along the line of sight. In either case, its
average value can be small.  In addition, for an individual maser the field
might fluctuate along the line of sight, reducing the total $RM$.

The interpretation of the linear polarization and RM data is thus currently
difficult and non-unique. Observations of more systems would be helpful and
may ultimately provide unique constraints on the thermal electron density
and/or magnetic field structure (e.g., reversals) in the nuclei of
(U)LIRGs.


\acknowledgments It is a pleasure to acknowledge Phil Perillat, who
performed the Mueller matrix calibration observations and reductions,
measured the antenna gain, and wrote the online data acquisition software
at Arecibo.  We thank Karen O'Neil, Amy Shelton, and Mark Clark for helping
us institute LSFS observing at the GBT.  This research benefited from
helpful discussions with Jeremy Darling, Vincent Fish, Peter Goldreich,
Bill Watson, Fred Lo, Moshe Elitzur, Willem Baan, and Loris Magnani.  TR
appreciates the technical guidance of grammarian Elena Cotto.  This
research was supported in part by NSF grant AST-0406987.  Support for this
work was also provided by the NSF to TR through awards GSSP 05-0001,
05-0004, and 06-0003, from the NRAO.  EQ was supported in part by NASA
grant NNG06GI68G and the David \& Lucile Packard Foundation.  This research
has made use of NASA's Astrophysics Data System Abstract Service and the
SIMBAD database, operated at CDS, Strasbourg, France.

{\it Facilities:} \facility{Arecibo}, \facility{GBT}


\bibliography{}


\clearpage

\def\arraystretch{1.2}
\tabletypesize{\normalsize}

\begin{deluxetable*}{lccccc}
\tablecolumns{6}
\tablewidth{0pt}
\tablecaption{IRAS F01417$+$1651 (III Zw 35) Gaussian Fit Parameters \label{tab:01417}}
\tabletypesize{\footnotesize}
\tablehead{
& \colhead{$S$} & \colhead{$\nu$} & \colhead{$\Delta\nu$} & \colhead{$v_{\odot}$} & \colhead{$B_{\parallel}$} \\
\colhead{Gaussian} & \colhead{(mJy)} & \colhead{(MHz)} & \colhead{(MHz)} & \colhead{(km s$^{-1}$)} & \colhead{(mG)} \\
\colhead{(1)} & \colhead{(2)} & \colhead{(3)} & \colhead{(4)} & \colhead{(5)} & \colhead{(6)}
}
\startdata
     0 \dotfill & $                        190.71 \pm   6.77\phantom{8}$ & $   1622.3743 \pm  0.0049$ & $   0.2041 \pm   0.0081$ & $   8312.6$ & $\phantom{-}     2.94 \pm   0.18$ \\
     1 \dotfill & $                        255.68 \pm  22.16           $ & $   1622.6881 \pm  0.0277$ & $   0.2786 \pm   0.0514$ & $   8253.0$ & $               -0.47 \pm   0.18$ \\
     2 \dotfill & $\phantom{8}              99.26 \pm   6.28\phantom{8}$ & $   1622.7237 \pm  0.0103$ & $   0.9694 \pm   0.0295$ & $   8246.2$ & $\phantom{-}     1.73 \pm   0.78$ \\
     3 \dotfill & $                        176.13 \pm  34.75           $ & $   1622.7604 \pm  0.0020$ & $   0.0905 \pm   0.0080$ & $   8239.3$ & $               -2.73 \pm   0.13$ \\
     4 \dotfill & $\phantom{8}\phantom{8}    1.17 \pm  59.50           $ & $   1622.8864 \pm  0.0142$ & $   0.1544 \pm   0.0239$ & $   8215.3$ & $               -3.59 \pm   0.26$
\enddata
\end{deluxetable*}

\clearpage

\begin{deluxetable*}{lccccc}
\tablecolumns{6}
\tablewidth{0pt}
\tablecaption{IRAS F10038$-$3338 (IC 2545) Gaussian Fit Parameters \label{tab:10038}}
\tabletypesize{\footnotesize}
\tablehead{
& \colhead{$S$} & \colhead{$\nu$} & \colhead{$\Delta\nu$} & \colhead{$v_{\odot}$} & \colhead{$B_{\parallel}$} \\
\colhead{Gaussian} & \colhead{(mJy)} & \colhead{(MHz)} & \colhead{(MHz)} & \colhead{(km s$^{-1}$)} & \colhead{(mG)} \\
\colhead{(1)} & \colhead{(2)} & \colhead{(3)} & \colhead{(4)} & \colhead{(5)} & \colhead{(6)}
}
\startdata
     0 \dotfill & $\phantom{8}              47.06 \pm   3.10\phantom{8}$ & $   1612.3870 \pm  0.0044$ & $   0.1505 \pm   0.0125$ & $  10221.0$ & $\phantom{-}\phantom{8}     1.58 \pm   0.75$ \\
     1 \dotfill & $                        179.73 \pm   5.15\phantom{8}$ & $   1612.8661 \pm  0.0032$ & $   0.1747 \pm   0.0077$ & $  10128.9$ & $           \phantom{ }    -1.76 \pm   0.26$ \\
     2 \dotfill & $\phantom{8}              83.85 \pm   5.50\phantom{8}$ & $   1613.0090 \pm  0.0119$ & $   0.9548 \pm   0.0405$ & $  10101.4$ & $                         -11.28 \pm   1.16$ \\
     3 \dotfill & $                        227.46 \pm  36.43           $ & $   1613.0362 \pm  0.0052$ & $   0.1222 \pm   0.0095$ & $  10096.2$ & $           \phantom{ }    -0.07 \pm   0.18$ \\
     4 \dotfill & $\phantom{8}\phantom{8}    2.85 \pm  14.04           $ & $   1613.1570 \pm  0.0089$ & $   0.1703 \pm   0.0157$ & $  10073.0$ & $           \phantom{ }    -0.13 \pm   0.16$ \\
     5 \dotfill & $\phantom{8}              97.85 \pm   4.75\phantom{8}$ & $   1613.3556 \pm  0.0035$ & $   0.1136 \pm   0.0075$ & $  10034.9$ & $\phantom{-}\phantom{8}     1.67 \pm   0.33$
\enddata
\end{deluxetable*}

\clearpage

\begin{deluxetable*}{lccccc}
\tablecolumns{6}
\tablewidth{0pt}
\tablecaption{IRAS F10173$+$0829 Gaussian Fit Parameters \label{tab:10173}}
\tabletypesize{\footnotesize}
\tablehead{
& \colhead{$S$} & \colhead{$\nu$} & \colhead{$\Delta\nu$} & \colhead{$v_{\odot}$} & \colhead{$B_{\parallel}$} \\
\colhead{Gaussian} & \colhead{(mJy)} & \colhead{(MHz)} & \colhead{(MHz)} & \colhead{(km s$^{-1}$)} & \colhead{(mG)} \\
\colhead{(1)} & \colhead{(2)} & \colhead{(3)} & \colhead{(4)} & \colhead{(5)} & \colhead{(6)}
}
\startdata
     0 \dotfill & $\phantom{8}              30.62 \pm   6.34$ & $   1589.2420 \pm  0.0037$ & $   0.0518 \pm   0.0113$ & $  14735.9$ & $\phantom{-}     2.34 \pm   2.50\phantom{8}$ \\
     1 \dotfill & $\phantom{8}              16.98 \pm   5.95$ & $   1589.2862 \pm  0.0041$ & $   0.0244 \pm   0.0106$ & $  14727.1$ & $\phantom{-}     4.19 \pm   3.42\phantom{8}$ \\
     2 \dotfill & $                        161.91 \pm   7.80$ & $   1589.3190 \pm  0.0024$ & $   0.1786 \pm   0.0050$ & $  14720.7$ & $\phantom{-}     0.25 \pm   0.89\phantom{8}$ \\
     3 \dotfill & $\phantom{8}              15.54 \pm   7.63$ & $   1589.3258 \pm  0.0073$ & $   0.0463 \pm   0.0240$ & $  14719.3$ & $               -2.95 \pm   5.15\phantom{8}$ \\
     4 \dotfill & $\phantom{8}\phantom{8}    0.42 \pm   1.58$ & $   1589.5315 \pm  0.0194$ & $   0.5661 \pm   0.0257$ & $  14678.6$ & $\phantom{-}     0.93 \pm   5.54\phantom{8}$ \\
     5 \dotfill & $\phantom{8}\phantom{8}    3.50 \pm   2.82$ & $   1589.6383 \pm  0.0154$ & $   0.0411 \pm   0.0412$ & $  14657.5$ & $               -2.90 \pm  17.78           $ \\
     6 \dotfill & $\phantom{8}              16.62 \pm   2.46$ & $   1589.8772 \pm  0.0072$ & $   0.1670 \pm   0.0279$ & $  14610.2$ & $\phantom{-}     0.80 \pm   7.57\phantom{8}$
\enddata
\end{deluxetable*}

\clearpage

\begin{deluxetable*}{lccccc}
\tablecolumns{6}
\tablewidth{0pt}
\tablecaption{IRAS F11506$-$3851 Gaussian Fit Parameters \label{tab:11506}}
\tabletypesize{\footnotesize}
\tablehead{
& \colhead{$S$} & \colhead{$\nu$} & \colhead{$\Delta\nu$} & \colhead{$v_{\odot}$} & \colhead{$B_{\parallel}$} \\
\colhead{Gaussian} & \colhead{(mJy)} & \colhead{(MHz)} & \colhead{(MHz)} & \colhead{(km s$^{-1}$)} & \colhead{(mG)} \\
\colhead{(1)} & \colhead{(2)} & \colhead{(3)} & \colhead{(4)} & \colhead{(5)} & \colhead{(6)}
}
\startdata
     0 \dotfill & $\phantom{8}              64.73 \pm  20.75           $ & $   1650.0094 \pm  0.0428$ & $   0.1910 \pm   0.0456$ & $   3152.3$ & $     1.21 \pm   0.27$ \\
     1 \dotfill & $                        102.55 \pm  26.92           $ & $   1650.1891 \pm  0.0240$ & $   0.1810 \pm   0.1018$ & $   3119.3$ & $     0.45 \pm   0.20$ \\
     2 \dotfill & $\phantom{8}              86.37 \pm  83.80           $ & $   1650.3232 \pm  0.0231$ & $   0.1358 \pm   0.0467$ & $   3094.7$ & $     0.36 \pm   0.17$ \\
     3 \dotfill & $                        116.30 \pm  24.97           $ & $   1650.4685 \pm  0.0257$ & $   0.2693 \pm   0.0976$ & $   3068.0$ & $     0.73 \pm   0.20$ \\
     4 \dotfill & $\phantom{8}\phantom{8}    0.31 \pm   7.25\phantom{8}$ & $   1650.7843 \pm  0.1171$ & $   0.3616 \pm   0.2594$ & $   3010.1$ & $     0.68 \pm   0.75$ \\
     5 \dotfill & $\phantom{8}              23.79 \pm   7.46\phantom{8}$ & $   1651.0731 \pm  0.0102$ & $   0.1260 \pm   0.0368$ & $   2957.1$ & $     1.03 \pm   0.50$
\enddata
\end{deluxetable*}

\clearpage

\begin{deluxetable*}{lccccc}
\tablecolumns{6}
\tablewidth{0pt}
\tablecaption{IRAS F12032$+$1707 Gaussian Fit Parameters \label{tab:12032}}
\tabletypesize{\footnotesize}
\tablehead{
& \colhead{$S$} & \colhead{$\nu$} & \colhead{$\Delta\nu$} & \colhead{$v_{\odot}$} & \colhead{$B_{\parallel}$} \\
\colhead{Gaussian} & \colhead{(mJy)} & \colhead{(MHz)} & \colhead{(MHz)} & \colhead{(km s$^{-1}$)} & \colhead{(mG)} \\
\colhead{(1)} & \colhead{(2)} & \colhead{(3)} & \colhead{(4)} & \colhead{(5)} & \colhead{(6)}
}
\startdata
     0 \dotfill & $\phantom{8}    4.31 \pm   0.54$ & $   1367.1002 \pm  0.0169$ & $   0.4082 \pm   0.0370$ & $  65844.0$ & $           \phantom{ }    -6.79 \pm   7.94\phantom{8}$ \\
     1 \dotfill & $\phantom{8}    3.95 \pm   0.96$ & $   1367.6739 \pm  0.0457$ & $   0.7915 \pm   0.1793$ & $  65690.6$ & $\phantom{-}               30.42 \pm  12.43           $ \\
     2 \dotfill & $\phantom{8}    8.42 \pm   0.99$ & $   1368.8016 \pm  0.0094$ & $   0.6264 \pm   0.0448$ & $  65389.5$ & $\phantom{-}\phantom{8}     6.91 \pm   4.86\phantom{8}$ \\
     3 \dotfill & $\phantom{8}    3.00 \pm   0.34$ & $   1369.5922 \pm  0.0071$ & $   0.1460 \pm   0.0207$ & $  65178.7$ & $           \phantom{ }    -1.05 \pm   6.48\phantom{8}$ \\
     4 \dotfill & $\phantom{8}    0.21 \pm   0.36$ & $   1369.6903 \pm  0.0452$ & $   2.6434 \pm   0.1574$ & $  65152.6$ & $           \phantom{ }    -3.25 \pm   3.98\phantom{8}$ \\
     5 \dotfill & $\phantom{8}    3.23 \pm   0.54$ & $   1369.9216 \pm  0.0036$ & $   0.0450 \pm   0.0089$ & $  65091.0$ & $\phantom{-}\phantom{8}     3.10 \pm   3.41\phantom{8}$ \\
     6 \dotfill & $\phantom{8}    6.67 \pm   0.34$ & $   1370.7194 \pm  0.0064$ & $   0.2954 \pm   0.0165$ & $  64878.6$ & $\phantom{-}               11.64 \pm   4.31\phantom{8}$ \\
     7 \dotfill & $              21.67 \pm   0.65$ & $   1371.2516 \pm  0.0030$ & $   0.4769 \pm   0.0115$ & $  64737.0$ & $\phantom{-}               10.90 \pm   1.72\phantom{8}$ \\
     8 \dotfill & $              16.99 \pm   0.46$ & $   1371.3316 \pm  0.0010$ & $   0.0849 \pm   0.0029$ & $  64715.8$ & $\phantom{-}               17.92 \pm   0.89\phantom{8}$ \\
     9 \dotfill & $              16.19 \pm   3.38$ & $   1372.1576 \pm  0.0568$ & $   0.4872 \pm   0.0531$ & $  64496.4$ & $\phantom{-}\phantom{8}     1.78 \pm   2.48\phantom{8}$ \\
    10 \dotfill & $              25.02 \pm   4.51$ & $   1372.3299 \pm  0.0027$ & $   0.2795 \pm   0.0165$ & $  64450.6$ & $           \phantom{ }    -1.45 \pm   1.12\phantom{8}$ \\
    11 \dotfill & $\phantom{8}    9.56 \pm   0.51$ & $   1372.7780 \pm  0.0412$ & $   0.7315 \pm   0.0588$ & $  64331.7$ & $                         -11.69 \pm   4.98\phantom{8}$ \\
    12 \dotfill & $\phantom{8}    2.02 \pm   0.21$ & $   1373.8371 \pm  0.0179$ & $   0.3807 \pm   0.0500$ & $  64051.0$ & $\phantom{-}               20.61 \pm  15.49           $
\enddata
\end{deluxetable*}

\clearpage

\begin{deluxetable*}{lccccc}
\tablecolumns{6}
\tablewidth{0pt}
\tablecaption{IRAS F12112$+$0305 Gaussian Fit Parameters \label{tab:12112}}
\tabletypesize{\footnotesize}
\tablehead{
& \colhead{$S$} & \colhead{$\nu$} & \colhead{$\Delta\nu$} & \colhead{$v_{\odot}$} & \colhead{$B_{\parallel}$} \\
\colhead{Gaussian} & \colhead{(mJy)} & \colhead{(MHz)} & \colhead{(MHz)} & \colhead{(km s$^{-1}$)} & \colhead{(mG)} \\
\colhead{(1)} & \colhead{(2)} & \colhead{(3)} & \colhead{(4)} & \colhead{(5)} & \colhead{(6)}
}
\startdata
     0 \dotfill & $              16.19 \pm   0.75$ & $   1554.1822 \pm  0.0013$ & $   0.1467 \pm   0.0054$ & $  21831.1$ & $               -0.20 \pm   1.99$ \\
     1 \dotfill & $              58.16 \pm   0.51$ & $   1554.2767 \pm  0.0048$ & $   0.5188 \pm   0.0098$ & $  21811.6$ & $\phantom{-}     0.25 \pm   1.17$ \\
     2 \dotfill & $              48.63 \pm   0.97$ & $   1554.5634 \pm  0.0007$ & $   0.1893 \pm   0.0028$ & $  21752.3$ & $\phantom{-}     1.02 \pm   0.77$ \\
     3 \dotfill & $              13.83 \pm   0.59$ & $   1554.5641 \pm  0.0006$ & $   0.0309 \pm   0.0017$ & $  21752.1$ & $\phantom{-}     0.27 \pm   1.03$ \\
     4 \dotfill & $\phantom{8}    0.32 \pm   0.34$ & $   1554.8544 \pm  0.0037$ & $   0.4359 \pm   0.0063$ & $  21692.1$ & $\phantom{-}     3.09 \pm   1.92$
\enddata
\end{deluxetable*}

\clearpage

\begin{deluxetable*}{lccccc}
\tablecolumns{6}
\tablewidth{0pt}
\tablecaption{IRAS F14070$+$0525 Gaussian Fit Parameters \label{tab:14070}}
\tabletypesize{\footnotesize}
\tablehead{
& \colhead{$S$} & \colhead{$\nu$} & \colhead{$\Delta\nu$} & \colhead{$v_{\odot}$} & \colhead{$B_{\parallel}$} \\
\colhead{Gaussian} & \colhead{(mJy)} & \colhead{(MHz)} & \colhead{(MHz)} & \colhead{(km s$^{-1}$)} & \colhead{(mG)} \\
\colhead{(1)} & \colhead{(2)} & \colhead{(3)} & \colhead{(4)} & \colhead{(5)} & \colhead{(6)}
}
\startdata
     0 \dotfill & $\phantom{8}    3.50 \pm   0.22$ & $   1315.2359 \pm  0.0105$ & $   0.4088 \pm   0.0317$ & $  80262.3$ & $                         -10.42 \pm  10.49           $ \\
     1 \dotfill & $\phantom{8}    8.95 \pm   0.19$ & $   1316.0759 \pm  0.0715$ & $   2.1625 \pm   0.0861$ & $  80019.7$ & $\phantom{-}               26.65 \pm   9.57\phantom{8}$ \\
     2 \dotfill & $\phantom{8}    5.70 \pm   0.30$ & $   1316.2886 \pm  0.0047$ & $   0.2040 \pm   0.0134$ & $  79958.4$ & $\phantom{-}\phantom{8}     1.76 \pm   4.65\phantom{8}$ \\
     3 \dotfill & $              11.26 \pm   0.68$ & $   1316.9061 \pm  0.0100$ & $   0.6771 \pm   0.0302$ & $  79780.3$ & $           \phantom{ }    -3.31 \pm   4.67\phantom{8}$ \\
     4 \dotfill & $\phantom{8}    0.08 \pm   0.50$ & $   1317.5519 \pm  0.0118$ & $   0.4878 \pm   0.0271$ & $  79594.2$ & $           \phantom{ }    -3.16 \pm   5.59\phantom{8}$ \\
     5 \dotfill & $\phantom{8}    2.78 \pm   0.26$ & $   1318.8802 \pm  0.0106$ & $   0.2664 \pm   0.0307$ & $  79212.1$ & $\phantom{-}               12.72 \pm  10.82           $ \\
     6 \dotfill & $              10.76 \pm   0.16$ & $   1319.1605 \pm  0.0113$ & $   1.2737 \pm   0.0199$ & $  79131.6$ & $                         -12.92 \pm   6.16\phantom{8}$
\enddata
\end{deluxetable*}

\clearpage

\begin{deluxetable*}{lccccc}
\tablecolumns{6}
\tablewidth{0pt}
\tablecaption{IRAS F15327$+$2340 (Arp 220) Gaussian Fit Parameters \label{tab:15327}}
\tabletypesize{\footnotesize}
\tablehead{
& \colhead{$S$} & \colhead{$\nu$} & \colhead{$\Delta\nu$} & \colhead{$v_{\odot}$} & \colhead{$B_{\parallel}$} \\
\colhead{Gaussian} & \colhead{(mJy)} & \colhead{(MHz)} & \colhead{(MHz)} & \colhead{(km s$^{-1}$)} & \colhead{(mG)} \\
\colhead{(1)} & \colhead{(2)} & \colhead{(3)} & \colhead{(4)} & \colhead{(5)} & \colhead{(6)}
}
\startdata
     0 \dotfill & $\phantom{8}              14.52 \pm   0.98$ & $   1637.1424 \pm  0.0008$ & $   0.0243 \pm   0.0020$ & $   5533.2$ & $               -4.78 \pm   0.53$ \\
     1 \dotfill & $\phantom{8}              10.92 \pm   0.61$ & $   1637.3105 \pm  0.0018$ & $   0.0709 \pm   0.0050$ & $   5501.9$ & $               -0.11 \pm   1.21$ \\
     2 \dotfill & $\phantom{8}              11.80 \pm   1.02$ & $   1637.5074 \pm  0.0009$ & $   0.0231 \pm   0.0024$ & $   5465.2$ & $               -2.78 \pm   0.64$ \\
     3 \dotfill & $\phantom{8}              10.12 \pm   1.00$ & $   1637.5736 \pm  0.0011$ & $   0.0241 \pm   0.0029$ & $   5452.9$ & $\phantom{-}     7.77 \pm   0.76$ \\
     4 \dotfill & $\phantom{8}\phantom{8}    0.90 \pm   0.77$ & $   1637.7198 \pm  0.0002$ & $   0.0590 \pm   0.0006$ & $   5425.6$ & $               -2.78 \pm   0.13$ \\
     5 \dotfill & $\phantom{8}              51.49 \pm   1.06$ & $   1637.8723 \pm  0.0005$ & $   0.0621 \pm   0.0016$ & $   5397.2$ & $\phantom{-}     0.33 \pm   0.25$ \\
     6 \dotfill & $                        324.77 \pm   3.96$ & $   1637.8916 \pm  0.0010$ & $   0.3362 \pm   0.0028$ & $   5393.6$ & $\phantom{-}     0.26 \pm   0.12$ \\
     7 \dotfill & $\phantom{8}              97.84 \pm   2.60$ & $   1638.0196 \pm  0.0008$ & $   0.0800 \pm   0.0023$ & $   5369.7$ & $               -0.15 \pm   0.18$ \\
     8 \dotfill & $                        293.07 \pm   4.06$ & $   1638.0313 \pm  0.0011$ & $   1.0064 \pm   0.0075$ & $   5367.6$ & $\phantom{-}     0.14 \pm   0.21$ \\
     9 \dotfill & $                        386.20 \pm   5.88$ & $   1638.1189 \pm  0.0007$ & $   0.0882 \pm   0.0012$ & $   5351.2$ & $               -0.76 \pm   0.06$ \\
    10 \dotfill & $\phantom{8}              81.13 \pm   2.84$ & $   1638.1375 \pm  0.0002$ & $   0.0292 \pm   0.0008$ & $   5347.8$ & $               -0.24 \pm   0.11$ \\
    11 \dotfill & $                        237.99 \pm   4.75$ & $   1638.2098 \pm  0.0011$ & $   0.1039 \pm   0.0024$ & $   5334.3$ & $\phantom{-}     0.66 \pm   0.10$ \\
    12 \dotfill & $\phantom{8}              46.95 \pm   2.66$ & $   1638.3468 \pm  0.0051$ & $   0.1894 \pm   0.0103$ & $   5308.8$ & $               -1.03 \pm   0.59$ \\
    13 \dotfill & $\phantom{8}              18.22 \pm   1.02$ & $   1638.4199 \pm  0.0008$ & $   0.0342 \pm   0.0024$ & $   5295.2$ & $\phantom{-}     0.20 \pm   0.50$ \\
    14 \dotfill & $\phantom{8}              19.79 \pm   0.78$ & $   1638.6066 \pm  0.0011$ & $   0.0769 \pm   0.0036$ & $   5260.4$ & $\phantom{-}     0.22 \pm   0.69$ \\
    15 \dotfill & $\phantom{8}\phantom{8}    7.62 \pm   0.60$ & $   1638.8445 \pm  0.0033$ & $   0.1038 \pm   0.0103$ & $   5216.1$ & $\phantom{-}     1.78 \pm   2.10$ \\
    16 \dotfill & $\phantom{8}              11.23 \pm   0.82$ & $   1638.9714 \pm  0.0014$ & $   0.0367 \pm   0.0034$ & $   5192.5$ & $\phantom{-}     1.42 \pm   0.86$ \\
    17 \dotfill & $\phantom{8}              13.18 \pm   0.57$ & $   1639.0663 \pm  0.0018$ & $   0.0908 \pm   0.0051$ & $   5174.9$ & $               -0.46 \pm   1.15$
\enddata
\end{deluxetable*}

\clearpage

\clearpage

\clearpage

\clearpage

\clearpage

\clearpage

\end{document}